\documentclass[lettersize,journal]{IEEEtran}
\usepackage{amsmath,amsfonts,amsthm,amssymb,bbm}
\usepackage{algorithm}
\usepackage{algpseudocode}
\usepackage{array}
\usepackage[caption=false,font=tiny,labelfont=sf,textfont=sf]{subfig}
\usepackage{textcomp}
\usepackage{stfloats}
\usepackage{url}
\usepackage{verbatim}
\usepackage{graphicx}
\usepackage{bm}
\usepackage{subfig}
\usepackage{booktabs}
\usepackage{ulem}
\usepackage{color}

\usepackage{makecell}

\def\BibTeX{{\rm B\kern-.05em{\sc i\kern-.025em b}\kern-.08em
    T\kern-.1667em\lower.7ex\hbox{E}\kern-.125emX}}
\usepackage{balance}
\renewcommand{\j}{\mathbf{j}}
\renewcommand{\vec}[1]{\ensuremath{\boldsymbol{#1}}}
\newcommand{\abs}[1]{\left| #1 \right|}
\newcommand{\norm}[1]{\left\|{#1}\right\|}

\newcommand{\T}{\mathrm{T}}

\newtheorem{theorem}{Theorem}
\newtheorem{assumption}{Assumption}
\newtheorem{lemma}{Lemma}
\newtheorem{definition}{Definition}

\begin{document}
\title{\huge DMRA: An Adaptive Line Spectrum Estimation Method through Dynamical Multi-Resolution of Atoms}

\author{Mingguang Han, Yi Zeng, Xiaoguang Li, and Tiejun Li
\thanks{M. Han, Y. Zeng, and T. Li are with Laboratory of Mathematics and Applied Mathematics, School of Mathematical Sciences,  Center for Machine Learning Research, Peking University, Beijing 100871, P.R. China}
\thanks{Email: mingguanghan@stu.pku.edu.cn (M. Han),  zengyi0427@pku.edu.cn (Y. Zeng), tieli@pku.edu.cn (T. Li)}
\thanks{X. Li is with MOE-LCSM, School of Mathematics and Statistics, Hunan Normal University, Changsha, Hunan 410081, P. R. China}
\thanks{Email: lixiaoguang@hunnu.edu.cn (X. Li)}
\thanks{Corresponding author: Xiaoguang Li, Tiejun Li}
}

\markboth{Journal of \LaTeX\ Class Files,~Vol.~18, No.~9, September~2024}%
{How to Use the IEEEtran \LaTeX \ Templates}

\maketitle

\begin{abstract}
We proposed a novel dense line spectrum super-resolution algorithm, the DMRA, that leverages dynamical multi-resolution of atoms technique to address the limitation of traditional compressed sensing methods when handling dense point-source signals. The algorithm utilizes a smooth $\tanh$ relaxation function to replace the $\ell_0$ norm, promoting sparsity and jointly estimating the frequency atoms and complex gains. To reduce computational complexity and improve frequency estimation accuracy, a two-stage strategy was further introduced to dynamically adjust the number of the optimized degrees of freedom. The strategy first increases candidate frequencies through local refinement, then applies a sparse selector to eliminate insignificant frequencies, thereby adaptively adjusting the degrees of freedom to improve estimation accuracy. Theoretical analysis were provided to validate the proposed method for multi-parameter estimations.  Computational results demonstrated that this algorithm achieves good super-resolution performance in various practical scenarios and outperforms the state-of-the-art methods in terms of frequency estimation accuracy and computational efficiency.
\end{abstract}


\section{Introduction}
The line spectrum estimation (LSE) problem is prevalent in communication engineering and physical sciences, with applications in areas such as direction of arrival (DOA) estimation \cite{Krim1996,Malioutov2005}, spectral analysis \cite{Stoica2005}, geophysical exploration \cite{Sacchi1998}, speech signal processing \cite{Eskimez2019}, and channel estimation in wireless communications \cite{Bajwa2010}, etc. In these applications, the spectrum of the measured signal consists of multiple discrete frequencies distributed over a continuous domain, with a minimum separation distance $\Delta_{\min}$ that may be smaller than the Rayleigh limit. This close-spacing frequency nature makes the LSE  a  challenging problem.


In recent years, the challenge of super-resolution for line spectral signals has gained much attention across various disciplines. In classical super-resolution algorithms, subspace-based methods such as the MUSIC algorithm \cite{Zoltowski1993,Liao2016}, the ESPRIT algorithm \cite{Roy1989}, and the Matrix Pencil method \cite{Hua1990} assume a known model order and perform eigen-decomposition on the data covariance matrix. These algorithms perform well with large sample size and high signal-to-noise ratio (SNR), but degrade with medium or low SNR. Another approach to LSE is the maximum likelihood (ML) technique \cite{Ziskind1988}, which usually relies on an accurate model order, typically determined by minimizing information criteria \cite{Wax1985,Stoica2004}. However, under small sample size and low SNR, these criteria often yield inaccurate model orders.

The theory of sparse representation in compressed sensing offers new insights to the LSE problem. If the parameters to be estimated are sparse or compressible under a certain dictionary, these parameters can be reconstructed with high probability under suitable conditions. Classical compressed sensing methods include greedy algorithms \cite{Tropp2007,Blumensath2009}, convex optimization algorithms \cite{Tibshirani1996,Figueiredo2007,Beck2009}, Sparse Bayesian Learning algorithms \cite{Wipf2004,Ji2008}, and non-convex optimization algorithms \cite{Chartrand2007,Foucart2009}, etc. By discretizing the continuous interval into a finite set of frequency grid points and using the discrete Fourier transform, the nonlinear LSE problem can be transformed into a linear  inverse problem with prior sparsity information \cite{Tropp2010}.

In general, the LSE problem deals with unknown frequencies of discrete point sources. Frequencies in a continuous domain may not align with pre-specified grids, leading to energy leakage and reducing LSE accuracy \cite{Chi2011}. To address this basis mismatch issue, some off-grid super-resolution algorithms are proposed, e.g., the off-grid sparse Bayesian inference \cite{Yang2012}, and parameterized dictionary refinement strategy \cite{Hu2012}, etc. While these algorithms mitigate the basis mismatch issue, their super-resolution performance for closely spaced sources is suboptimal. Methods like atomic norm minimization \cite{Tang2013,Bhaskar2013,Tang2015}, interpreted as Beurling-Lasso (BLASSO) estimators with an infinitely refined grid \cite{Koulouri2020}, also face limitations. Theoretically, stable recovery is only feasible when frequencies are well separated, and the computational cost remains high \cite{Candes2014,Suliman2021,Chi2020}.

To achieve super-resolution of closely spaced point sources, algorithms must simultaneously overcome the basis mismatch problem and ensure sparse solutions. The reweighted $\ell_1$ algorithm, proposed by Candès et al. \cite{candes2008}, iteratively minimizes $\sum_{n=1}^N\mathrm{log}(|x_n| + \epsilon)$ to promote sparsity, and has been shown to efficiently recover closely-sparse signals \cite{Ahmad2015}. Subsequent theoretical analyses \cite{Chartrand2008,Daubechies2010,lai2013improved,Chen2018} confirm the effectiveness of reweighted $\ell_1$ and related $\ell_2$ algorithms. Dynamic dictionary parameter learning versions \cite{Fang2016,Ye2018} further improve these algorithms, though the pre-set dictionaries still face basis mismatch issue. The Newtonized orthogonal matching pursuit (NOMP) algorithm \cite{Mamandipoor2016} was also proposed to address basis mismatch by extending OMP to the continuous parameter space through off-grid optimization. Though successful, its computational cost is still relatively high.

By integrating the effective iterative re-weighting idea, we propose an adaptive line spectrum estimation method through Dynamical Multi-Resolution of Atoms (DMRA) technique with carefully designed sparsity selectors. The algorithm deals with the basis mismatch problem while keeping the solution sparse. The main contributions of this paper are as follows:
\begin{enumerate}{}{}
\item{\textbf{Super-Resolution Relaxation Optimization}: We relax the $\ell_0$ norm penalty by the $\mathrm{tanh}(|x|^2/\epsilon)$ function, achieving super-resolution of multiple dense point source signals in the continuous domain through joint optimization of dictionary parameters and signal gains. Compared to the SURE-IR algorithm \cite{Fang2016} which uses the $\mathrm{log}$ relaxation function, our method shows superior super-resolution performance in the continuous domain and is also applicable to other dense parameter dictionary models.}
\item{\textbf{Dynamical Multi-Resolution}: Our algorithm dynamically adjusts the degrees of freedom (DoF) based on a two-stage procedure, i.e. the on-grid and off-grid estimators. In on-grid stage, the candidate frequencies are gradually narrowed down through dynamical multi-resolution grids; while an off-grid stage is used to refine the frequencies and reduce energy leakage. The algorithm ensures data fitting with the minimum necessary DoF, significantly enhancing the sparsity and accuracy of frequency estimation.}
\item{\textbf{Theoretical Analysis}: We conduct a theoretical analysis of the tanh-sum function minimization problem, demonstrating that as $\epsilon\to 0$,  the global minimum of the tanh-sum minimization approaches the true solution, providing a partial theoretical validation of the considered relaxation strategy.}
\end{enumerate}

\textbf{Outline}:
The organization of the rest of this paper is as follows: In Section~\ref{sec:sec2},  we describe the problem setup of  line spectral super-resolution and formulate it as a parameter estimation problem for dynamic sparse dictionaries. In Section~\ref{sec:sec3}, we develop the DMRA algorithm for the line spectrum estimation with dynamic grids and varying DoF. The theoretical analysis of the optimization problem in the noiseless scenario is performed in Section~\ref{sec:sec4}. The numerical results are demonstrated in Section~\ref{sec:sec5}, and we finally make the conclusion in Section~\ref{sec:sec6}.

\textbf{Notation}:
In this paper, the matrices and vectors are denoted by uppercase and lowercase bold letters, respectively.  $\mathcal{R}\{ x\}$ represents the real part of a complex number $x$. The superscripts $(\cdot)^*$, $(\cdot)^H$ and $(\cdot)^\T$ denote conjugate, conjugate transpose and transpose, respectively. The symbol $\lceil \cdot \rceil$ denotes the ceiling function, and $\|\cdot\|_2$ the $\ell_2$-norm. We denote $|\cdot|$ the absolute value or modulus of a real/complex number, and $\#|\cdot|$ the number of elements in a set.  The Kronecker product is denoted by $\otimes$, the Khatri-Rao product by $\circledcirc$, and the Hadamard product by $\odot$. $\mathbbm{1}_N$ denotes an $N$-dimensional vector with all elements equal to 1. $\mathcal{V}[X]$ represents the vectorization of matrix $X$. $\text{diag}(\bm{x})$ denotes a diagonal matrix with vector $\bm{x}$ on its diagonal. $\Delta_{\mathrm{dft}} = 1/M$ represents the standard resolution in discrete Fourier transform (DFT). For a given frequency $\omega$, we denote its corresponding frequency atom  as $\vec{a}(\omega):= [1, e^{-\j 2\pi\omega}, \cdots, e^{-\j 2\pi(M-1)\omega}]^\T$, where $\j$ is the imaginary unit with the definition $\j^2=-1$.  Given a set of frequencies $\Omega=\{\omega_1, \omega_2,\dots, \omega_K\}$, we always assume they are sorted in a proper order and write them in a vector $\vec{\omega}=\vec{\omega}(\Omega)=[\omega_1, \omega_2,\dots, \omega_K]^\T$. We may sometimes treat the vector $\vec{\omega}$ as a set if this does not lead to ambiguity. The dictionary matrix corresponding to a frequency vector $\vec{\omega}$ is denoted as $A(\vec{\omega})=[\vec{a}(\omega_1),\dots, \vec{a}(\omega_K)]$.

\section{Problem Description}\label{sec:sec2}

Consider a frequency-sparse signal consists of $S$ discrete point sources/atoms
\begin{equation}\label{eq:01}
  x_m = \sum_{k=1}^S h_k^r e^{-\j 2\pi (m-1)\omega_k^r},\quad m=1,2,\dots, M,
\end{equation}
where $\omega_k^r$ is the frequency of the $k$-th atom, and $h_k^r \in \mathbb{C}$ denotes the corresponding complex gain.  We use the superscript `$r$' to indicate that it is for ground truth. Without loss of generality, we assume the frequency $\{\omega_k^r\}_{k=1}^S$ is defined in the periodic interval  $\mathbb{T}=[0,1)$. The corresponding atomic measure of the signal is $\mu(\vec{\omega}^r) = \sum_{k=1}^{S}h^r_k\delta(\omega-\omega_k^r)$. Considering the additive circularly-symmetric Gaussian white noise (AGN), the received signal is
\begin{equation}\label{eq:02}
  \vec{y}=\sum_{k=1}^{S}h_k^r\vec{a}(\omega_k^r) + \vec{\eta}, \quad \vec{\eta}\sim\mathcal{CN}(0, \sigma^2\mathbb{I}_M).
\end{equation}
Taking the convention of dictionary matrix, the system can also be expressed in a matrix form
\begin{equation}\label{eq:03}
  \vec{y}=A(\vec{\omega}^r)\vec{h}^r + \vec{\eta},
\end{equation}
where the dictionary matrix $A(\vec{\omega}^r)\in\mathbb{C}^{M\times S}$, and $\vec{h}^r=[h_1^r,\ldots,h^r_S]^\T\in \mathbb{C}^{S}$.

We assume the signal has a clustered structure \cite{Batenkov2021, Li2021}. The $S$ atoms form $K$ ($K<S$) clusters. In each cluster, the atoms are closely distributed while the distance between the atoms in different clusters is relatively large. Specifically, the corresponding atomic measure of the signal has the form
\begin{equation}\label{eq:04}
  \mu(\vec{\omega}^r) = \sum_{k=1}^{S}h_k^r\delta(\omega-\omega_k^r) = \sum_{k=1}^{K}\sum_{j=1}^{n_k}h_{j,k}^r\delta(\omega-\omega_{j,k}^r),
\end{equation}
where $\omega_{j,k}^r$ denotes the position of the $j$-th atom in cluster $\Lambda_k$, and $n_1+n_2+\dots +n_K=S$. In cluster $\Lambda_k$, the distance between neighbouring atoms is generally small such that $\mathrm{dist}(\omega_{j,k}^r,\omega_{j+1,k}^r)<\Delta_{\mathrm{dft}}$, $\forall j,k$; while the distance between different clusters $\text{dist}(\Lambda_m,\Lambda_n)> \alpha\Delta_{\mathrm{dft}}$ for some $\alpha>1$.

The aim of LSE is to estimate the model order $S$, the frequency support $\vec{\omega}$, and the complex gains $\{h_j\}_{j=1}^S$ of the mixed discrete point sources. The super-resolution of LSE is typically a non-linear optimization problem even the model order $S$ is known. The capability of super-resolution is determined by the minimum separation distance, which is defined as bellow:

\begin{definition}\label{Definition:1}
(\textbf{Minimum Separation Distance}): Given a vector $\bm{\omega} = [\omega_1, \omega_2, \cdots, \omega_S]\in \mathbb{T}^S$ with $S$ distinct atoms, the minimum separation distance is defined as
\[
\Delta_{\min} := \min_{i \neq j} | \omega_i - \omega_j |_{\mathbb{T}} \quad \text{where} \quad | \omega |_{\mathbb{T}} := \min_{n \in \mathbb{Z}} |\omega - n|.
\]
\end{definition}
When the minimum separation distance is small, the LSE problem becomes difficult. It is shown in \cite{Chi2020} that signal sources can be theoretically guaranteed to be recovered by the atomic norm minimization only when the minimum separation distance $\Delta_{\min}\geq 2.52\Delta_{\mathrm{dft}}$.  Different mathematical models of sparsity also affect the super-resolution capabilities of algorithms \cite{Batenkov2021, Li2021}.  Traditional compressed sensing approaches discretize the continuous parameter space $\mathbb{T}$ into a finite grid set, using this predefined dictionary to transform the compressed sensing problem into a linear inverse problem. However,  the mismatch between the true frequency points and the predefined dictionary often deteriorates the performance of compressed sensing algorithms.

To address this issue, we consider the joint recovery of sparse signal frequencies and complex gains with an unknown parameter dictionary. Given the noise level $\delta$, we aim to find the minimal combination of atoms from the parameter set $\vec{\omega} \in \mathbb{T}^N$ based on the observations $\bm{y} \in \mathbb{C}^M$. The parameter $\vec{\omega}$ is to be dynamically optimized. The sparse signal recovery problem can be modeled as the following $\ell_0$ norm minimization problem:
\begin{equation}\label{eq:05}
(\mathcal{P}) \quad \min_{\bm{h} \in \mathcal{S}_{\bm{\omega}}, \bm{\omega} \in \mathbb{T}^N} \|\bm{h}\|_0 \quad \text{s.t.} \quad \| \bm{y} - A(\bm{\omega}) \bm{h} \|_2 \leq \delta,
\end{equation}
here, $\mathcal{S}_{\bm{\omega}} = \{ \bm{h} \in \mathbb{C}^N : \| \bm{y} - A(\bm{\omega}) \bm{h} \|_2 \leq \delta \}$, $\|\bm{h}\|_0 = \#|\{n : h_n \neq 0\}|$, and $\delta$ represents the noise level.

However, even with fixed parameters $\bm{\omega}$, the problem $(\mathcal{P})$ is a combinational optimization problem. We have to relax it to a smooth optimization problem that promotes sparsity in the solution. In this paper, we replace $\norm{\vec{h}}_0$ by a smooth function $g(\cdot)$ (more precisely, the sum of $g(h_n)$ for $\vec{h}=[h_1,\ldots,h_N]^\T$) and solve a surrogate problem
\begin{equation}\label{eq:06}
\min_{\bm{h}\in \mathcal{S}(\bm{\omega}),\bm{\omega}\in \mathbb{T}^N} \sum_{n=1}^N g(h_n) \quad \text{s.t.} \quad \|\bm{y} - A(\bm{\omega})\bm{h}\|_2 \leq \delta.
\end{equation}
Here, $g(\cdot)$ is a relaxation function with similar properties of the function $\|x\|_0$. When $g(x)=\abs{x}$, the problem \eqref{eq:06} is convex. The sparse solution of this $\ell_1$ minimization problem can be efficiently obtained by various compressed sensing algorithms. However, it is limited by the restricted isometry property (RIP) conditions, requiring more observations and providing weaker super-resolution for dense line spectra in practice.

\begin{figure}[!t]
\centering
\includegraphics[width=2.36in]{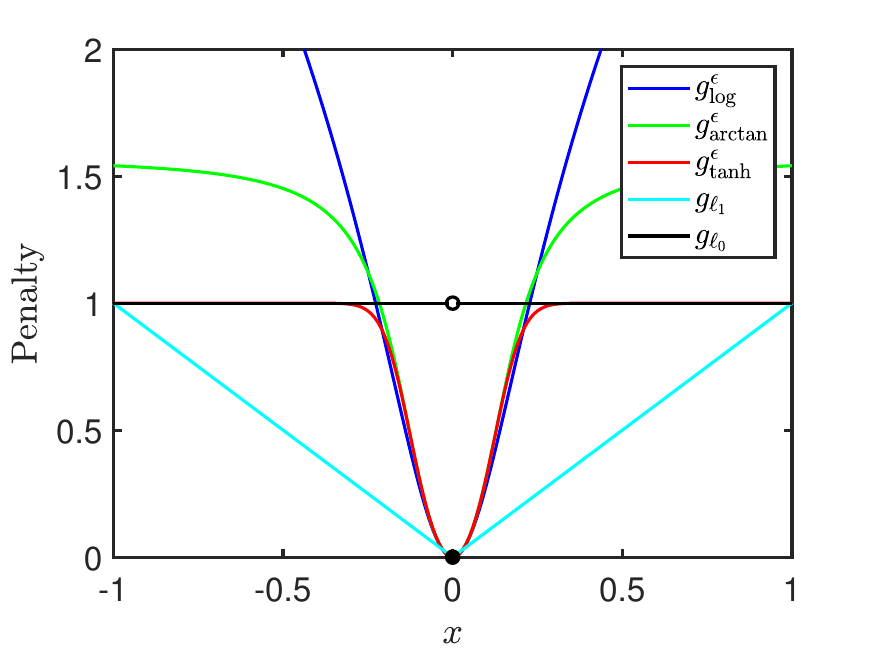}
\caption{Schematics of four relaxation functions $g_{\mathrm{log}}^{\epsilon}(x), g_{\mathrm{atan}}^{\epsilon}(x), g_{\mathrm{tanh}}^{\epsilon}(x)$, and $g_{\ell_1}(x)$ versus the original  $\ell_0$ norm function: $g_{\ell_0}(x)=\|x\|_0$.}
\label{fig:1}
\end{figure}

Motivated by \cite{Fang2016} and the intuition to relax the $\ell_0$ norm, we compare several smooth non-convex approximations to $g_{\ell_0}(\vec{x}):=\|\vec{x}\|_0$, such as $g_{\mathrm{log}}^{\epsilon}(x) = \log(1 + |x|^2/\epsilon)$, $g_{\mathrm{atan}}^{\epsilon}(x) = \mathrm{atan}(|x|^2/\epsilon)$, and $g_{\mathrm{tanh}}^{\epsilon}(x) = \mathrm{tanh}(|x|^2/\epsilon)$. By adjusting the relaxation parameter $\epsilon$, these functions control the strength and smoothness of the penalty, approximating the $\ell_0$ norm to achieve sparsity. By setting $\epsilon=0.03$, Fig.~\ref{fig:1} shows a schematics of these relaxation functions versus the original $\ell_0$ norm. We can find that all these three non-convex relaxation functions approximate the $\ell_0$ norm better than the $\ell_1$ norm. Among them, $g_{\mathrm{tanh}}^{\epsilon}$ is the closest approximations. As the tradeoff, we need to pay the cost to perform a non-convex optimization for the relaxed problem, provided we can design efficient super-resolution algorithms. This is the starting point of our study.

Therefore, in this paper, we use the hyperbolic tangent function to promote sparsity instead of the $\ell_0$ function, replacing $(\mathcal{P})$ with the following problem $(\mathcal{P}_1)$:
\begin{equation}\label{eq:07}
\begin{aligned}
(\mathcal{P}_1) \quad \min_{\bm{h} \in \mathcal{S}_{\bm{\omega}}, \bm{\omega} \in \mathbb{T}^N}  &\mathcal{L}^{\epsilon}(\bm{h}) = \sum_{n=1}^N \mathrm{tanh} \left( {|h_{n}|^2}/{\epsilon} \right) \\
 \text{s.t.}\ & \|\bm{y} - A(\bm{\omega}) \bm{h}\|_2 \leq \delta,
\end{aligned}
\end{equation}
where $\epsilon>0$ is a small constant ensuring the well-defined nature of the $\mathrm{tanh}$ function. We solve this problem by reformulating it to the following unconstrained optimization problem:
\begin{equation}
\begin{aligned}\label{eq:08}
(\mathcal{P}_1')\quad \min_{\bm{\omega}, \bm{h}} \mathcal{L}(\bm{\omega}, \bm{h}) = &\|\bm{y} - A(\bm{\omega}) \bm{h}\|_2^2 \\
& + \lambda \sum_{n=1}^N \mathrm{tanh} \left( {|h_{n}|^2}/{\epsilon} \right)
\end{aligned}
\end{equation}
for some constant $\lambda>0$. In Section~\ref{sec:sec4}, we will show that in the noiseless scenario, as $\epsilon$ approaches 0, the solution to the relaxed problem $(\mathcal{P}_1)$ tends to the solution of problem $(\mathcal{P})$. In numerical experiments, the parameter $\epsilon$ is initialized as a relatively large value and gradually decreases to $0$. This may keep the computing from prematurely converging to a local optima.

\section{An Adaptive Line Spectrum Estimation Method through Dynamical Multi-Resolution of Atoms}\label{sec:sec3}

When minimizing $(\mathcal{P}_1')$, the size of the dictionary $N$ governs the computational cost. To reduce the computational complexity and improve the accuracy, we propose the DMRA, an adaptive two stage dynamical multi-resolution of atoms technique with a preprocessing step, to achieve the goal. Before the two-stage LSE optimization, we first perform a preprocessing  by ignoring the relaxation function and applying a DFT estimator on canonical grid to roughly estimate the complex gains and make a frequency initialization based on this estimate. Then, an on-grid estimation (Stage 1) is proposed on a locally refined grid to get a more accurate $\vec{h}$ and $\vec{\omega}$. Finally, we take an off-grid estimation (Stage 2) in which we solve a fully nonlinear optimization problem to refine the value of frequencies and update the complex gains correspondingly. At the end of each stage, a sparsity selector is followed to adjust the size of the dictionary.

\subsection{Preprocessing: Frequency Initialization}

In the preprocessing step, we initialize the dictionary as $\vec{\omega}^{(i)}=[0, 1/M,\ldots, (M-1)/M]^\T$, which makes the matrix $A(\bm{\omega}^{(i)})$ the standard DFT matrix.  Ignoring the relaxation function, the initial MMSE estimator of $\vec{h}$ is $\hat{\bm{h}}^{(i)} = A^{H}(\bm{\omega}^{(i)})\bm{y}/M$. $\hat{\bm{h}}^{(i)}$ is then passed through a hard-thresholding \textit{Sparsity Selector A} to select the frequencies that meet the thresholding criteria.The design of \textit{Sparsity Selector A} is as follows:

We set the hard threshold $\psi_a$ by combining the error comes from AGN and the energy leakage effect:
\begin{equation}\label{eq:09}
\psi_{a} = \sigma^2 \frac{\log M}{M} + \gamma_{a} \frac{E_{\text{tot}}}{S^{\text{pri}}},
\end{equation}
where the first term in $\psi_{a}$  originates from \cite{Bhaskar2013, Tang2015}, and the second term reflects the interference energy leakage of the complex sinusoids due to basis mismatch. Here, $\gamma_{a}$ is a constant related to the energy decay of the complex sinusoids, with a value less than 1, used to balance the proportion of energy leakage and noise. $S^{\text{pri}}$ is the prior sparsity. The total signal energy $E_{\text{tot}}$ is estimated by $ E_{\text{tot}} = \max(\|\bm{y}\|_2^2/M - \sigma^2, 0)$. 

Then the energy components of $\vec{\bm{h}}^{(i)}$ that are greater than $\psi_a$ and the corresponding frequencies are selected:
\begin{equation}\label{eq:11}
\vec{\omega}^{(0)} = \{ \omega_k \in \vec{\omega}^{(i)} : |\hat{h}_k^{(i)}|^2 \geq \psi_{a},\ k=1,2,\ldots,M\}.
\end{equation}

In the preprocessing step, we just roughly select the frequencies on the canonical DFT grids by ignoring the frequencies with weak energy which may correspond to noise. The frequency initialization only requires a FFT operation, whose computational cost is $O(M\log M)$.

\subsection{Stage 1: On-Grid Estimation}

 In this stage, we solve problem $(\mathcal{P}_1')$ on a locally refined multi-resolution grid to enhance sparsity. For some integer $\gamma\geq 0$, the locally refined grid is set to be
 \begin{align}\label{eq:12}
 \tilde{\vec{\omega}}=\Big\{ \omega_k^{(0)}\pm \frac{m}{2\gamma+1} & \Delta_{\mathrm{dft}},\ m=0,1,\dots,\gamma; \nonumber \\
& \omega_k^{(0)}\in\vec{\omega}^{(0)},\ k=1,2,\dots N_0\Big\},
 \end{align}
where $N_0:=\#|\vec{\omega}^{(0)}|$ and $\gamma$ is also referred to as the grid refinement factor. Because the problem $(\mathcal{P}_1')$ is nonlinear, we solve it by a Majorization-Minimization (MM) iteration strategy \cite{candes2008}. From the concaveness of $\tanh$ function on the positive side of real axis, we have
\begin{equation}\label{eq:13}
  \tanh(x) \leq \tanh(x_0) + (1-\tanh^2(x_0))(x - x_0)
\end{equation}
for any $x, x_0\in\mathbb{R}^+$. We adopt the following heuristic strategy to perform a successive optimization of a surrogate objective function $Q(\cdot)$ with respect to $\vec{h}$. Denote $\tilde{\vec{h}}^k$ the complex gain $\vec{h}$ and $\tilde{\vec{\omega}}^k$ the identified frequency point at the $k$-th iteration step. Take
\begin{equation}\label{eq:14}
\begin{aligned}
  Q(\vec{h}|&\tilde{\vec{h}}^k) = \norm{\vec{y} - A(\tilde{\vec{\omega}}^k)\vec{h}}^2_2+\lambda \sum_{n=1}^N \tanh \left(|\tilde{h}^k_{n}|^2/\epsilon \right)\\
  & + \frac{\lambda}{\epsilon}\sum_{n=1}^{N}\Big(1-\tanh^2(|\tilde{h}^k_n|^2 / \epsilon)\Big)\big(\abs{h_n}^2-|\tilde{h}_n^k|^2\big).
\end{aligned}
\end{equation}
We have that $Q(\vec{h}|\tilde{\vec{h}}^k)$ is a majorization of $\mathcal{L}(\tilde{\vec{\omega}}^k,\vec{h})$ in the sense that $\mathcal{L}(\tilde{\vec{\omega}}^k,\vec{h})\leq Q(\vec{h}|\tilde{\vec{h}}^k)$ and $\mathcal{L}(\tilde{\vec{\omega}}^k,\tilde{\vec{h}}^k) = Q(\tilde{\vec{h}}^k|\tilde{\vec{h}}^k)$. The surrogate function $Q(\vec{h}|\tilde{\vec{h}}^k)$ is a quadratic form of $\vec{h}$ and its global minimum point  can be explicitly found to be
\begin{equation}\label{eq:15}
\tilde{\bm{h}}^{k+1} = \big (A^H(\tilde{\vec{\omega}}^k)A(\tilde{\vec{\omega}}^k) + \frac{\lambda}{\epsilon} \mathrm{diag} \left( \bm{\mathrm{w}}^k \right) \big)^{-1}A^H(\tilde{\vec{\omega}}^k)\bm{y},
\end{equation}
where $\bm{\mathrm{w}}^k = 1 - \bm{\vartheta}_{h} \odot \bm{\vartheta}_{h}, \bm{\vartheta}_{h} = \tanh (\tilde{\bm{h}}^k \odot (\tilde{\bm{h}}^k)^*/\epsilon )$.

\begin{algorithm}[!t]
\caption{On-Grid Estimation (Stage 1)}
\label{alg:On_Grid_Estimation}
\begin{algorithmic}[1]
\State \textbf{Input:} Signal $\bm{y}$, refinement factor $\gamma$, prior sparsity $S^{\text{pri}}$
\State \textbf{Output:} Estimated frequencies $\bm{\omega}^{(1)}=\text{final }\tilde{\vec{\omega}}$
\State \textbf{Initialize:} $\gamma_{b}$, dictionary matrix $A(\tilde{\bm{\omega}})$ with $\vec{\omega}^{(0)}$ and $\gamma$
\State $G \leftarrow A^H(\tilde{\vec{\omega}})A(\tilde{\vec{\omega}})$, $\bm{\widetilde{y}} \leftarrow A^H(\tilde{\vec{\omega}})\bm{y}$
\State $N \leftarrow \#|\tilde{\bm{\omega}}|$
\While{$N > S^{\text{pri}}$}
    \State Compute $\tilde{\bm{h}}$ using \eqref{eq:15}
    \State Compute $\psi_{b} \leftarrow \frac{\gamma_b}{N}\|\tilde{\bm{h}}\|_2^2$
    \State Update $\tilde{\vec{\omega}} \leftarrow \bm{\omega}(\tilde{h}_i^* \cdot \tilde{h}_i \geq \psi_b)$, $N_p \leftarrow \#|\tilde{\vec{\omega}}|$
    \If{$N_p = N$} \textbf{break}
    \State $N \leftarrow N_p$, Update $G$ and $\bm{\widetilde{y}}$
    \EndIf
\EndWhile
\end{algorithmic}
\end{algorithm}

To further reduce the computational cost of the MM iteration, we apply the \textit{Sparsity Selector B} after each MM iteration step.
In \textit{Sparsity Selector B}, we remove the frequency points in $\tilde{\vec{\omega}}^k$ whose corresponding amplitude is less than the threshold $\psi_b=\gamma_b\|\tilde{\vec{h}}^{k+1}\|_2^2/N$, where $0<\gamma_b<1$. That is, we update the frequency as
\begin{equation*}
\tilde{\vec{\omega}}^{k+1}=\Big\{\tilde{\omega}_n^k\in\tilde{\vec{\omega}}^k: |\tilde{h}_n^{k+1}|^2>\psi_b, n=1,\ldots,\#|\tilde{\vec{\omega}}^{k}|\Big\}.
\end{equation*}
 After this selection, the $(k+1)$-th step of MM iteration has fewer variables and can be solved faster. The MM iteration stops until the number of selected frequencies reaches the prior sparsity $S^{\text{pri}}$. The complete algorithm is described in Algorithm~\ref{alg:On_Grid_Estimation}.

Since the basis used in each iteration is a subset of the basis in the previous step, $A(\tilde{\vec{\omega}}^k)$ is a sub-matrix of $A(\tilde{\vec{\omega}}^{0})$, which is the initial matrix refined from Eq.~(\ref{eq:12}). So we only need to compute the Gram matrix $G = A^H(\tilde{\vec{\omega}}^{0})A(\tilde{\vec{\omega}}^{0})$ and $\widetilde{\bm{y}} = A^H(\tilde{\vec{\omega}}^{0})\bm{y}$ once. The computational cost of determining $G \in \mathbb{C}^{N \times N}$ is $\mathcal{O}(N^2M)$. The main cost in each iteration is computing \eqref{eq:15}. To estimate the overall cost in the on-grid estimation, we assume that after each sparsity selection, only $\lceil \kappa N \rceil$ atoms remain if there are $N$ atoms in previous step, where $\kappa<1$ is a constant.  Then we have $A(\tilde{\vec{\omega}}^{0}) \in \mathbb{C}^{M\times N_1}$, and after the $k$-th iteration, $A(\tilde{\vec{\omega}}^k) \in \mathbb{C}^{(\lceil \kappa^{k-1} N_1\rceil)\times (\lceil \kappa^{k-1}N_1\rceil)}$, where $N_1=\#|\tilde{\vec{\omega}}^{0}|$. Assuming the total number of iterations is $R_1$, we get the total computational cost of the on-grid optimization stage as  $\mathcal{O}(N_1^3(1-\kappa^{3R_1})/(1-\kappa^3))$.

This MM iteration together with sparsity selection procedure can also be regarded as a reweighted algorithm \cite{candes2008}. Unlike traditional reweighted algorithms that impose finite weights on atoms, we eliminate the atoms  with small amplitudes at each iteration, which is equivalent to imposing an infinite weight on those atoms. Practically, this approach more effectively promotes sparsity in the solution. Our on-grid stage iteratively and adaptively reduces the DoF, searching for smaller and more precise subspaces, thereby reducing the size of the dictionary matrix and the complexity of the algorithm. This may enhance the accuracy and efficiency of the super-resolution of LSE.

\begin{algorithm}[!t]
\caption{Off-Grid Estimation (Stage 2)}
\label{alg:Off_Grid_Estimation}
\begin{algorithmic}[1]
\State \textbf{Input:} Observed signal $\bm{y}$, Threshold parameter $\gamma_{c}$, $\beta$, $T_v$
\State \textbf{Output:} Estimated frequencies $\bm{\omega}^{(2)}=\text{final }\hat{\bm{\omega}}$

\State \textbf{Initialize:} $\bm{\xi} \leftarrow [\bm{\nu}^\T, \bm{\omega}^\T, \bm{\phi}^\T]^\T$, $\hat{\bm{\omega}}^{(0)} \leftarrow \hat{\bm{\omega}}$, $N \leftarrow \#|\hat{\bm{\omega}}|$
\For{$i = 1$ to $\max\_\text{iterations}$}
    \State $\bm{\xi} \leftarrow [\bm{\nu}^\T, \hat{\bm{\omega}}^\T, \bm{\phi}^\T]^\T$
    \State Solve \eqref{eq:16} with BFGS optimization to obtain $\hat{\bm{\xi}}$
    \State Calculate residual $\bm{y}_r = \bm{y} - \Psi(\hat{\bm{\omega}}, \hat{\bm{\phi}})\hat{\vec{\nu}}$
    \If{$\|\mathcal{F}^{-1}{(\bm{y}_r)}\|_{\infty}^2 > T_v$}
        \State $\gamma_c \leftarrow 0.8\gamma_c$, \quad $\beta \leftarrow 0.8\beta$
    \Else
        \If{$|\hat{\bm{\omega}}| \leq N$}
            \State $N \leftarrow \#|\hat{\bm{\omega}}|$
        \EndIf
        \State $\gamma_c \leftarrow 1.1\gamma_c$, \quad $\beta \leftarrow 1.1\beta$
    \EndIf
    \State $\hat{\bm{\omega}} \leftarrow [\hat{\bm{\omega}}^{(0)}, \hat{\bm{\omega}}]$
    \State Use Algorithm~\ref{alg:sparse_selector_c} to update $\hat{\bm{\omega}}$
\EndFor
\end{algorithmic}
\end{algorithm}

\subsection{Stage 2: Off-Grid Estimation}

 In the preprocessing and on-grid estimation stage, the problem $(\mathcal{P}_1')$ is solved on a predefined grid, though refined, it is still not sufficient to achieve good enough accuracy. In  the off-grid stage, we will  solve the fully nonlinear problem $(\mathcal{P}_1')$ to further refine the estimation of frequencies. At the same time, a \textit{Sparsity Selector C} is carefully designed to keep the sparsity of the reconstructed signal.

Denote the final dictionary selected by the on-grid stage as $\vec{\omega}^{(1)}$, we are going to solve the following problem
\begin{equation}\label{eq:16}
\min_{\vec{\xi}} \mathcal{L}(\vec{\xi}) = \|\bm{y} - \Psi(\bm{\omega}, \bm{\phi})\bm{\nu}\|_2^2 + \lambda \sum_{n = 1}^N \tanh\left({\nu_n^2}/{\epsilon}\right),
\end{equation}
where the complex gain $\vec{h}$ is decomposed into the amplitude $\vec{\nu}$ and angle $\vec{\phi}$ as $\vec{h}=\vec{\nu}\odot (e^{-\j \vec{\phi}})$ with $\nu_n = |h_n|$, $\phi_n = -\text{arg} (h_n)$, $[\Psi]_{mn} = e^{-\j (2\pi m \omega_n + \phi_n)}$. The target variable $\bm{\xi} = [\bm{\nu}^\T, \vec{\omega}^\T, \bm{\phi}^\T]^\T$ is a real vector. For such nonlinear continuous optimization problem, we utilize the efficient quasi-Newton type methods, the BFGS method, in our off-grid stage \cite{Noc2006}.

The overall off-grid estimation stage (Stage 2) is described in Algorithm \ref{alg:Off_Grid_Estimation}. The basic idea is to first apply a continuous optimization (e.g., the BFGS method) to get an estimator $\hat{\vec{\xi}}$. Then we apply a sparsity selector (\textit{Sparsity Selector C}) to further refine the frequency points. After that, we perform an accuracy checking step (i.e., the CFAR stopping criterion below). If the accuracy checking test is passed, the whole process is done; otherwise, we need to update some thresholds and redo the optimization. Below let us describe the concrete procedures in detail.

Denote the estimated solution of \eqref{eq:16} after the first round optimization as $\hat{\vec{\xi}}=[\hat{\nu}^\T, \hat{\vec{\omega}}^\T, \hat{\vec{\phi}}^\T]^\T$. We are going to retain as few DoFs as possible while ensuring that the fidelity residual reaches the noise level. This task  is done by \textit{Sparsity Selector C}. First we merge the frequencies close enough into a single one. Then the frequency point whose complex gain is less than a threshold $\psi_c$ will be removed. Specifically, given a sorted sequence of frequencies $\hat{\vec{\omega}}$, if the distance between adjacent frequencies $\Delta_{k,k+1} = |\hat{\omega}_{k+1} - \hat{\omega}_k|<\beta$, then $\hat{\omega}_k$ and $\hat{\omega}_{k+1}$ are merged into a new frequency $\hat{\omega}_k' = \tau \hat{\omega}_k + (1 - \tau) \hat{\omega}_{k+1}$. The weight $\tau = e_k/(e_k + e_{k+1})$ is calculated based on the energy magnitudes, where $e_k = |\hat{h}_k|^2$ and $e_{k+1} = |\hat{h}_{k+1}|^2$. The merged energy is updated as $e_k' = \tau e_k + (1 - \tau) e_{k+1}$. This process is repeated to refine all adjacent frequencies until all pairs of frequencies satisfy the distance threshold. After this merging step, the frequency atoms with energy less than the threshold $\psi_c=\|\hat{\vec{h}}\|^2_2\gamma_c/N$ are removed from the dictionary. The algorithm for \textit{Sparsity Selector C} is summarized in Algorithm \ref{alg:sparse_selector_c}.

In \textit{Sparsity Selector C}, the parameters $\beta$ and $\gamma_c$ should be carefully chosen. If  $\beta$ or $\gamma_c$ is too small, we may keep too many insignificant frequencies so  the computational complexity cannot be reduced; on the other hand, if $\beta$ or $\gamma_c$ is too large, we may eliminate too much information that the reconstructed signal cannot meet the noise level. To properly choose $\beta$ and $\gamma_c$, we adopt the Constant False Alarm Rate (CFAR) stopping criterion \cite{Mamandipoor2016,Han2019} to measure whether the reconstructed signal has met the noise level.  Under the hypothesis $\hat{\vec{\omega}}=\vec{\omega}^r$ and $\hat{\vec{h}}=\vec{h}^r$, the residual of the reconstructed signal satisfies $\mathcal{F}^{-1}(\bm{y}_r) :=\mathcal{F}^{-1}(\bm{y} - \Psi(\hat{\bm{\omega}}, \hat{\bm{\phi}})\hat{\bm{\nu}})=A^H(\vec{\omega}^{(i)})\vec{\eta}/M$. Due to the orthogonality of $A(\vec{\omega}^{(i)})$, $\tilde{\vec{\eta}}=A^H(\vec{\omega}^{(i)})\vec{\eta}/\sqrt{M}$ is still an AGN with covariance $\sigma^2\mathbb{I}$. It is straightforward to show that
\begin{equation}\label{eq:17}
\begin{aligned}
\mathbb{P}&\left\{\norm{\mathcal{F}^{-1}(\vec{y}_r)}_{\infty}^2<T_v\big|\hat{\vec{\omega}}=\vec{\omega}^r, \hat{\vec{h}}=\vec{h}^r\right\} \\ &=  \mathbb{P}\left\{\|\tilde{\bm{\eta}}\|_{\infty}^2 < MT_v \right\}  = \left(1 - e^{-\frac{M T_v}{\sigma^2}}\right)^M.
\end{aligned}
\end{equation}

\begin{algorithm}[!t]
\caption{\textit{Sparsity Selector \(C\)} in Off-Grid Estimation}
\label{alg:sparse_selector_c}
\begin{algorithmic}[1]
\State \textbf{Input:} Frequencies $\hat{\vec{\omega}}$, gain $\hat{\bm{h}}$, threshold parameter $\beta$, $\gamma_c$
\State \textbf{Output:} Estimated frequencies $\hat{\bm{\omega}}$
\State $N \leftarrow |\hat{\vec{\bm{\omega}}}|$, \quad $\bm{e} \leftarrow \hat{\bm{h}} \odot \hat{\bm{x}}$

\For{$k = 1 \to N-1$}
    \If{$|\hat{\omega}_{k+1} - \hat{\omega}_k| < \beta$}
        \State $\tau \leftarrow \frac{e_k}{e_k + e_{k+1}}$
        \State $\hat{\omega}_k \leftarrow \tau \hat{\omega}_k + (1 - \tau) \hat{\omega}_{k+1}$
        \State $e_k \leftarrow \tau e_k + (1-\tau)e_{k+1}$
        \State Remove $\hat{\omega}_{k+1}$ and $e_{k+1}$ from $\hat{\bm{\omega}}$ and $\bm{e}$
        \State $N \leftarrow N - 1$
    \EndIf
\EndFor

\State $\psi_c \leftarrow \gamma_c \frac{1}{N} \|\hat{\bm{h}}\|_2^2$
\State $\hat{\bm{\omega}} \leftarrow \{\hat{\omega}_i \in \hat{\bm{\omega}} : e_i > \psi_c\}$
\end{algorithmic}
\end{algorithm}

Let $T_v = \sigma^2(\log(M)+Z)/M$. We have $(1 - e^{-M T_v/\sigma^2})^M=\left(1 - M^{-1}e^{-Z}\right)^M$. Since $\left(1 - M^{-1} t\right)^M\to e^{-t}$ as $M\to\infty$, we have $(1 - M^{-1} e^{-Z})^M \approx \exp\left(-\exp(-Z)\right)$ when $M$ is large. Thus we have
\[\mathbb{P}\left\{\norm{\mathcal{F}^{-1}(\vec{y}_r)}_{\infty}^2>T_v\big|\hat{\vec{h}}=\vec{h}^r \right\}\approx 1-\exp\big(-\exp(-Z)\big).\]
Given a small probability $P_{\text{fa}}$, we can ensure the probability
\begin{equation}\label{eq:18}
\mathbb{P}\left\{\norm{\mathcal{F}^{-1}(\vec{y}_r)}_{\infty}^2>T_v\big|\hat{\vec{\omega}}=\vec{\omega}^r, \hat{\vec{h}}=\vec{h}^r\right\}<P_{\text{fa}}
\end{equation}
by choosing $Z =\log(M) - \log(-\log(1-P_{\text{fa}}))$, namely $T_v=M^{-1}\sigma^2(\log(M) - \log(-\log(1-P_{\text{fa}})))$. Therefore, if we found $\norm{\mathcal{F}^{-1}(\vec{y}_r)}^2_{\infty}>MT_v$, we can reject the original hypothesis $\hat{\vec{\omega}}=\vec{\omega}^r$ and $\hat{\vec{h}}=\vec{h}^r$ with high probability, which means the reconstructed signal does not meet the noise level so we should shrink $\beta$ and $\gamma_c$. Otherwise, $\beta$ and $\gamma_c$ will be increased.

 Having modified $\beta$ and $\gamma_c$, the continuous optimization problem \eqref{eq:16} will be re-performed to further refine $\vec{\omega}$. The whole procedure will be repeated until some stopping criterion is satisfied. After each threshold adjustment, we use the initial frequency atoms of the current stage to expand the DoFs of the model and re-evaluate the corresponding unknowns that include the initial frequency atoms. Additionally, local grid refinement near the current baseline can be performed to further enhance the accuracy of the computation.

For clarity, we summarize the overall DMRA algorithm in Algorithm \ref{alg:DMRA}. The proposed algorithm is a low-complexity adaptive frequency selection algorithm. In preprocessing stage, the power spectrum filtering reduces the frequency search space by a relatively small computational cost. This allows the subsequent refinement of the grid within a subspace with much fewer DoFs compared to the entire space. The overall approach of Stages 1 and 2 is to use basis with more DoFs by local refinement, and then use a sparse selector to eliminate  insignificant frequencies, adaptively changing the number of DoFs to improve frequency estimation accuracy. During the discrete optimization phase, $G$ and $\widetilde{\bm{y}}$ only need to be calculated at the initial stage, and subsequent iterations do not require recalculation. Therefore, it reduces the computational load of the discrete optimization phase.

\begin{algorithm}[!t]
\caption{DMRA: Adaptive Line Spectrum Estimation Method through Dynamical Multi-Resolution of Atoms}
\label{alg:DMRA}
\begin{algorithmic}[!t]
\State \textbf{Input:} Signal $\bm{y}$, refinement factor $\gamma$, prior sparsity $S^{\text{pri}}$
\State \textbf{Output:} Estimated frequencies $\hat{\bm{\omega}}=\bm{\omega}^{(2)}$

\State \textbf{Initializing:} $\gamma_{a}$, $\gamma_{b}$, $\gamma_{c}$, $\beta$, $T_v$

\State \textbf{Preprocessing: Frequency Initialization}
\State Compute $\psi_{a}$ using \eqref{eq:09}
\State $\hat{\bm{h}}^{(i)} \leftarrow \mathcal{F}^{-1}(\bm{y})$
\State Select frequencies $\vec{\omega}^{(0)}$ using \eqref{eq:11}
\State \textbf{Stage 1: On-Grid Estimation}
\State Use Algorithm \ref{alg:On_Grid_Estimation} to estimate $\vec{\omega}^{(1)}=\tilde{\vec{\omega}}$
\State \textbf{Stage 2: Off-Grid Estimation}
\State Use Algorithms \ref{alg:Off_Grid_Estimation} to estimate $\bm{\omega}^{(2)}=\hat{\vec{\omega}}$
\end{algorithmic}
\end{algorithm}

In \cite{Mamandipoor2016}, the authors discussed the necessity of oversampling in the Newtonized step, which is also required in Stage 1 to further compress the subspace and find initial values. Since we use continuous optimization in Stage 2, the factor $\gamma$ does not need to be very large. An excessively large $\gamma$ will increase the computational load of the Off-Grid estimation. It is recommended to progressively refine the grid to achieve $2\gamma+1$ times instead of directly refining by $2\gamma+1$ times, which can further reduce the computational cost of the algorithm.

The selection of $\gamma_a, \gamma_b$ is relatively straightforward and typically set to a fixed small value, their selection is simple. In the Off-Grid stage of Algorithm \ref{alg:DMRA}, $\gamma_c$ and $\beta$ can be adaptively updated, with relatively large initial values. The thresholds $\psi_b$ and $\psi_c$ can be adaptively updated based on the current estimated coefficients. Numerical experiments indicate that these parameters are not sensitive to the algorithm's performance. The regularization parameter $\lambda$ is related to the noise and signal energy. When the noise level is unknown, the adaptive selection scheme for the regularization parameter from \cite{Fang2016} can be adopted. Similar to \cite{Fang2016}, gradually reducing $\epsilon$ during the optimization process can further increase the probability of finding the correct solution.

We remark that the DMRA algorithm can be easily extended to the multiple measurement vector (MMV) model \cite{Li2015} for line spectrum recovery, following steps similar to those used for the SMV model. In the numerical experiments, we applied this method to channel estimation and frequency domain extrapolation problems, utilizing the quasi-Newton method for continuous optimization. Details of the optimization process can be referred  to \cite{Noc2006}.

\section{Theoretical Analysis}\label{sec:sec4}

In this section, we will show in the noiseless scenario, the solution to the relaxed problem $(\mathcal{P}_1)$ is exactly the solution of $\ell_0$ minimization problem $(\mathcal{P}_0)$ under proper conditions.

We consider the true line spectrum signal $\bm{y}^r = A(\bm{\omega}^r)\bm{h}^r$, where the columns of $A(\bm{\omega}^r)$ are linearly independent and $M \geq 2S$. The true complex sinusoids is exactly the superposition of $S$ different complex exponential atoms. When $\delta = 0$, the problem $(\mathcal{P})$ can be written as $(\mathcal{P}_0)$:
\begin{equation}\label{eq:19}
(\mathcal{P}_0)\quad \min_{\bm{h} \in \mathcal{S}_{\bm{\omega}}, \bm{\omega} \in \mathbb{T}^N} \|\bm{h}\|_0 \quad \text{s.t.} \quad \bm{y} = A(\bm{\omega})\bm{h}.
\end{equation}

Problems $(\mathcal{P})$ and $(\mathcal{P}_0)$ aim to find the simplest explanation that fits the data. Theorem~\ref{theorem:1} below states that in the absence of noise, the global optimal solution of $(\mathcal{P}_0)$ is the true signal.

\begin{theorem}\label{theorem:1}
    Let $(\bm{\tilde{\omega}^r}, \bm{\tilde{h}^r})$ be the globally optimal solution of $(\mathcal{P}_0)$, satisfying $M \geq 2S$ and $S \geq 2$. Then, $(\bm{\tilde{\omega}^r}, \bm{\tilde{h}}^r)$ is unique and equal to the true solution $(\bm{\omega}^r, \bm{h}^r)$, with $\|\tilde{\bm{h}}^r\|_0 = S$.
\end{theorem}

The proof is provided in Appendix~A.

Problem $(\mathcal{P}_0)$ is a combinational optimization problem. We use a relaxation problem $(\mathcal{P}_\epsilon)$ to approximate $(\mathcal{P}_0)$:
\begin{equation}\label{eq:20}
\begin{aligned}
(\mathcal{P}_\epsilon)\quad \min_{\bm{h} \in \mathcal{S}_{\bm{\omega}}, \bm{\omega} \in \mathbb{T}^N} & \mathcal{L}^{\epsilon}(\bm{h}) = \sum_{n=1}^N \mathrm{tanh}\left(\frac{|h_n|^2}{\epsilon}\right) \\
& \text{s.t.} \quad \bm{y} = A(\bm{\omega})\bm{h}.
\end{aligned}
\end{equation}

Since the true signal is composed of $S$ different atoms, there is a constant $\tau>0$ such that $\min_{i\neq j}\abs{\omega^r_i-\omega^r_j}>\tau$. Denote $\vec{\omega}^\epsilon$ the globally optimal solution to problem $(\mathcal{P}_\epsilon)$, we make the following assumption on the separation of $\vec{\omega}^\epsilon$.

\begin{assumption}\label{assumption:1}
    The condition $\min_{i\neq j}\abs{\omega^\epsilon_i-\omega^\epsilon_j}>\tau$ holds for any $\epsilon>0$. Moreover, if $\vec{\omega}^r\cap\left(\cup_{\epsilon>0}\vec{\omega}^\epsilon\right)=\emptyset$, then $\inf_{\epsilon}\min_{i,j}\abs{\omega^r_i-\omega^\epsilon_j}>\tau$.
\end{assumption}

To show the equivalence between the problems $(\mathcal{P}_0)$ and $(\mathcal{P}_\epsilon)$, we need the following  lemma.

\begin{lemma}\label{lemma:1}
Let $M, S\in\mathbb{N}_{+}$, $M\geq2S$. $\vec{\omega}\in\mathbb{T}^S$. Let $\Phi_{M,S}(\bm{\omega})=\left[ e^{-\j 2\pi m \omega_n}\right]_{m=0:M-1}^{n=1:S}\in\mathbb{C}^{M\times S}$ be a complex Vandermonde matrix. Let $\Delta_{\min}(\bm{\omega})$ denote the minimum separation distance between atoms in $\bm{\omega}$, $\mu = \Delta_{\min}(\bm{\omega}) / \Delta_{\mathrm{dft}}$, and $\Theta_{M,S}(\bm{\omega}) := \min_{\|\omega_i - \omega_j\|_{\mathbb{T}} \geq \Delta_{\min}} \sigma_{\min}(\Phi_{M,S}(\bm{\omega}))$, where $\sigma_{\min}(\Phi)$ is the smallest singular value of matrix $\Phi$. Then there exists a constant $C$ such that
\[
\Theta_{M,S}(\bm{\omega}) \geq C \sqrt{M} \mu^{S-1}.
\]
\end{lemma}

The proof of Lemma \ref{lemma:1} can be found in Theorem 3 in \cite{Li2021} and Theorem 3.2 in \cite{Batenkov2020}.

Now we are ready to show that under Assumption \ref{assumption:1}, the problems $(\mathcal{P}_0)$ and $(\mathcal{P}_\epsilon)$ have the same solution when $\epsilon$ is small enough.

\begin{theorem}\label{theorem:2}
    Let $(\bm{\omega}^r, \bm{h}^r)$ be the globally optimal solution of problem $(\mathcal{P}_{0})$, and let $(\bm{\omega}^{\epsilon}, \bm{h}^{\epsilon})$ be the globally optimal solution of problem $(\mathcal{P}_{\epsilon})$, where $\bm{\omega}^{\epsilon}$ contains $P$ non-zero elements. Under Assumption 1, there exists a constant $C$ such that when
\begin{equation}\label{21}
\epsilon \leq \frac{C^2 M^{4S-2} \tau^{4S-2} \gamma_{\max}^2}{(P-S) \mathrm{ln}(1 + S)},
\end{equation}
we have $\mathcal{L}^{\epsilon}(\bm{h}^{\epsilon}) = \mathcal{L}^{\epsilon}(\bm{h}^r)$, meaning that the globally optimal solution of the relaxed problem $(\mathcal{P}_{\epsilon})$ is equivalent to the optimal solution of $(\mathcal{P}_{0})$.
\end{theorem}

The proof is provided in Appendix~B.

\section{Simulation Result}\label{sec:sec5}

In this section, we conduct extensive testing of the DMRA method across various scenarios with detailed evaluation criteria. We first test the single measurement vector (SMV) scenario. Subsequently, we use the 5G Toolbox of MATLAB to generate the CDL-C spatial-frequency channel and evaluate the performance of our method in channel estimation.

\subsection{SMV Scenario}

In the SMV scenario, we test the line spectral estimation problem for dense point sources. The signal is a mixture of $S$ discrete point sources from $K$ clusters, expressed as
\[
\bm{y} = \sum_{k=1}^K \sum_{l=1}^{n_k} h_{l,k}^r \bm{a}(\omega_{l,k}^r) + \bm{\eta} = \bm{y}^r + \bm{\eta},
\]
where $\bm{y}^r$ is the true signal in the observation domain, and $S = \sum_{k=1}^{K} n_k$.

\textbf{Evaluation Criteria.}
\begin{itemize}
    \item Reconstruction Signal-to-Noise Ratio (RSNR): RSNR measures the reconstruction accuracy of the observed signal. Given the reconstructed signal $\hat{\bm{y}}$, $\text{RSNR} := \mathbb{E} \left\{ 20 \log_{10} \left( \|\bm{y}^r\|_2/\|\hat{\bm{y}} - \bm{y}^r\|_2 \right) \right\}$.
    \item Super-Resolution Success Rate: Each true frequency $\omega_i^r$ has a neighborhood $N_{\omega_i} := \{ \omega: |\omega - \omega_i| < 0.15 \times \Delta_{\mathrm{dft}} \}$. If the estimated frequency is not within $N_{\omega_i}$, $\omega_i^r$ is considered lost; otherwise, it is successfully detected. If all frequencies are successfully detected and the mean squared error between estimated frequencies $\{\hat{\omega}_i\}_{i=1}^S$ and true frequencies $\{\omega_i^r\}_{i=1}^S$ is less than $3 \times 10^{-3}$, i.e., $\|\hat{\bm{\omega}} - \bm{\omega}^r\|_2 \leq 3 \times 10^{-3}$, the super-resolution recovery is considered as successful.
	\item Frequency Estimation Error: The accuracy is measured by the normalized mean squared error (NMSE), $\text{NMSE} := \mathbb{E} \left\{ \sum_{s=1}^S (\hat{\omega}_s - \omega_s^r)^2 / (S \Delta_{\mathrm{dft}}^2) \right\}$. If $\omega_i^r$ is not detected, $\hat{\omega}i = \omega_i^r + 0.3 \Delta{\mathrm{dft}}$ is used.
	\item Computation Time: Super-resolution performance is tested in MATLAB R2021b, comparing CPU time across algorithms.
\end{itemize}

\textbf{Simulation Setup.} We consider a signal comprised of 8 complex sinusoids of length $M = 100$. Four scenarios are defined based on different cluster counts and $\Delta_{\min}/\Delta_{\mathrm{dft}}$ ratios, as detailed in Table \ref{tab:1}. Scenarios 1, 2, and 3 are each tested 100 times at various SNR levels. In Scenario 4, we retain two clusters but vary the number of point sources, with 100 tests conducted per source count and SNR level. The cluster separation parameter is set to $\alpha = 10$. $\text{SNR}_{\text{norm}}$ denotes the normalized SNR, assuming equal SNR for all sinusoids. Gain $|h_{l,k}^r| = \sigma \sqrt{\text{SNR}_{\text{norm}}}$ and phase $\phi_{l,k}^r = \angle h_{l,k}^r$ are uniformly distributed in $[0,2\pi)$. Frequencies are randomly chosen from $[0,1)^S$, and only samples adhering to the scenario-specific cluster structure are retained.

\begin{table}[!t]
\centering
\caption{Tested Scenarios in the Evaluation}
\label{tab:1}
\begin{tabular}{cccc}
    \hline\hline
    Scenario & \makecell{Source Number $S$ \\ ($n_k$ sources per cluster)} & \makecell{$K$\\(Number of clusters)} & $\mu = \frac{\Delta_{\min}}{\Delta_{\mathrm{dft}}}$ \\
    \hline
    1 & $8 \; (n_1 = 3, n_2 = 2, n_3 = 3)$ & 3 & 0.5 \\
    2 & $8 \; (n_1 = 4, n_2 = 4)$ & 2 & 0.5 \\
    3 & $8 \; (n_1 = 8)$ & 1 & 0.8 \\
    4 & $4:2:16 \; (n_1 = \frac{S}{2}, n_2 = \frac{S}{2})$ & 2 & 0.8 \\
    \hline\hline
\end{tabular}
\end{table}

We conduct numerical experiments to evaluate the performance of the proposed DMRA method. The initial parameters are set as $\gamma_{a} = 0.05$, $\gamma_{b} = 0.2$, $\gamma_{c} = 0.8$, and $\beta = 0.5\Delta_{\mathrm{dft}}$. Assume the sparsity prior $S^{\text{pri}}$ is 20 for Scenarios 1-3 and $2S$ for Scenario 4. The initial regularization parameter $\lambda = \sigma^2/(E_{\text{tot}}/S^{\text{pri}})$, where $E_{\text{tot}}$ is estimated as described in Section \ref{sec:sec3}, is adaptively updated using the current signal energy estimate. The grid refinement factor is $\gamma = 5$.

\textbf{Benchmark Algorithms.} We compare the proposed algorithm with several effective line spectrum estimation methods.

For on-grid methods, we test the sparse convex optimization method (MFISTA) \cite{Wei2022}, iterative reweighted algorithms (SURE-IR \cite{Fang2016} and IRLS \cite{Chartrand2008}), and the Root-MUSIC algorithm \cite{Zoltowski1993}. We also include the AST method with ADMM solver \cite{Hansen2019, Chi2020} and NOMP, known for fast convergence and low computational complexity \cite{Mamandipoor2016}.

In MFISTA and IRLS, we replace the optimization part with the BFGS quasi-Newton algorithm, denoted as Off-MFISTA and Off-IR$\ell_2$. All Off-MFISTA, Off-IR$\ell_2$, SURE-IR, and NOMP algorithms require preset initial grid points, $\bm{\omega}^{(i)} = [0, \cdots, (2\gamma+1) M-1]^\T/((2\gamma+1) M)$. Based on empirical evidence, Off-MFISTA, Off-IR$\ell_2$, and SURE-IR require denser grids, so we set $\gamma = 5$. For NOMP, we set $\gamma = 2$. The MUSIC algorithm searches for frequencies by traversing spectrum peaks on a discrete grid, with points set as $\bm{\omega}^{(i)} = [0, \cdots, 100M-1]^\T/(100M)$.

We then test the performance of different algorithms across various scenarios. Each scenario uses 100 samples with different random seeds. The final results are the average of these 100 samples.

\noindent\paragraph{\textbf{RSNRs and Success Rates}}

\begin{figure}[!t]
    \centering
    \subfloat[\label{fig:2.1}]{
    \includegraphics[width=2.36in]{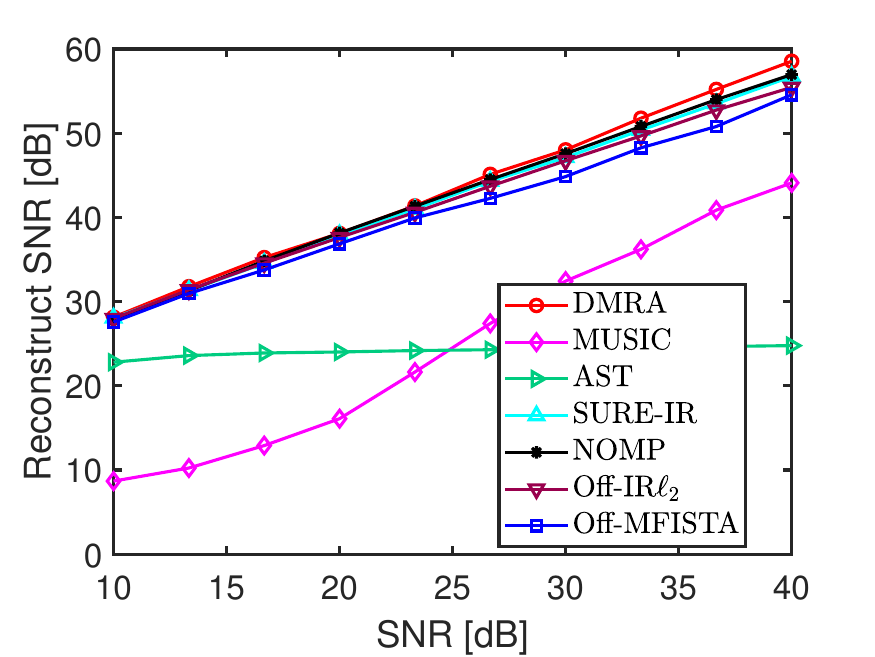}}
    \hspace{0.000\textwidth}
    \subfloat[\label{fig:2.2}]{
    \includegraphics[width=2.36in]{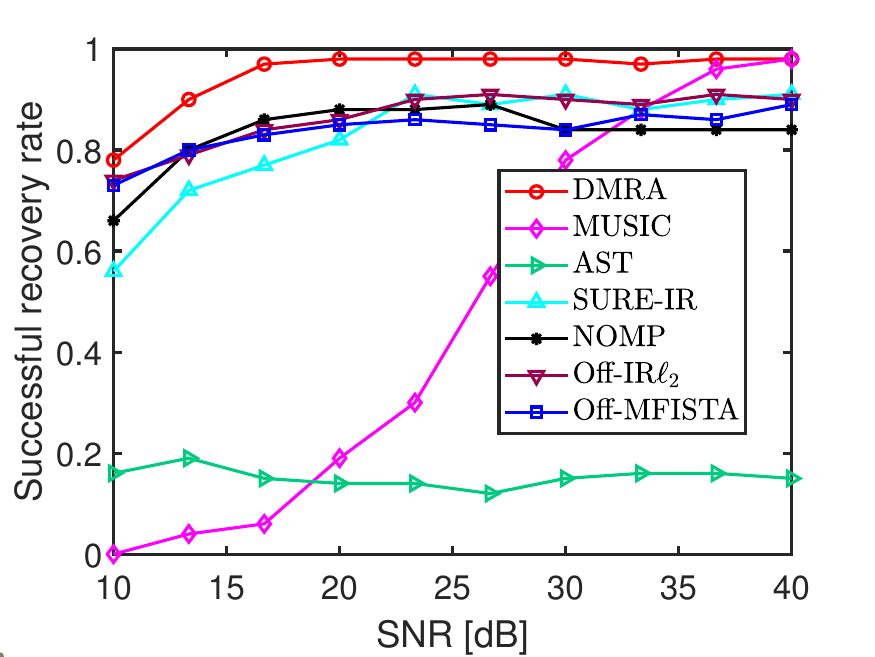}}
    \caption{The RSNR and success rates of various algorithms for different normalized SNR in Scenario 1. The result is the average of 100 independent tests. (a): The average RSNR of different algorithms. (b): The success rates of different algorithms.}
    \label{fig:2}
\end{figure}

We first test Scenario 1, where point sources are distributed in 3 clusters. A valid sample is generated if $\Delta_{\min} \geq 0.5\Delta_{\mathrm{dft}}$ and the sources conform to the cluster structure of Scenario 1 in Table \ref{tab:1}. Fig. \ref{fig:2} shows the average RSNR and success rates of different algorithms as a function of normalized SNR.

As shown in Fig.~\ref{fig:2}, the DMRA algorithm outperforms the others across various SNRs, particularly in high SNR regions, where its RSNR and success rates are significantly higher. For Scenario 1, DMRA and the refined algorithms (Off-IR$\ell_2$, Off-MFISTA) maintain high success rates. While SURE-IR, Off-IR$\ell_2$, and Off-MFISTA improve with increasing SNR, they remain slightly inferior to DMRA overall.

\begin{figure}[!t]
    \centering
    \subfloat[\label{fig:3.1}]{
    \includegraphics[width=2.36in]{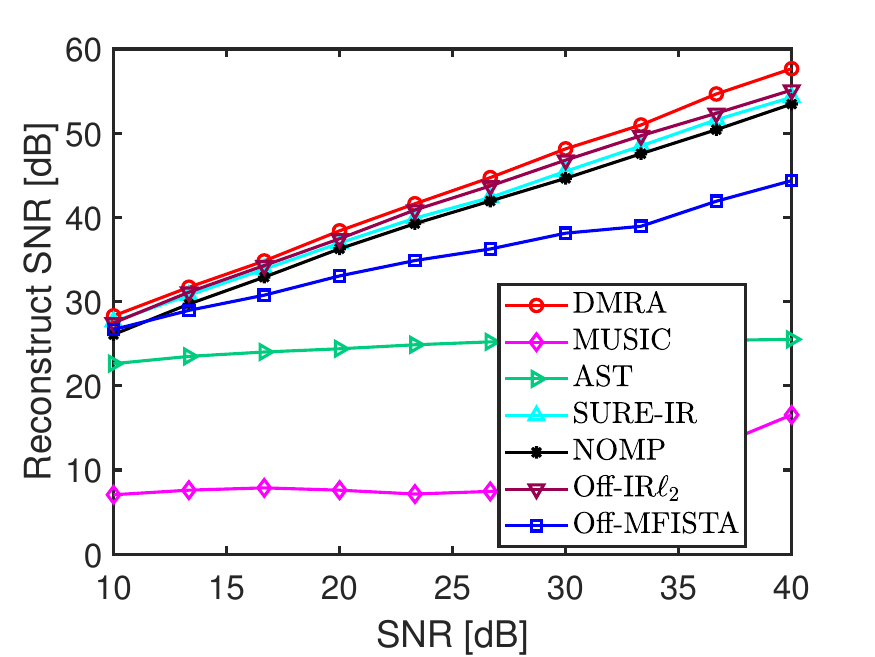}}
    \hspace{0.00\textwidth}
    \subfloat[\label{fig:3.2}]{
    \includegraphics[width=2.36in]{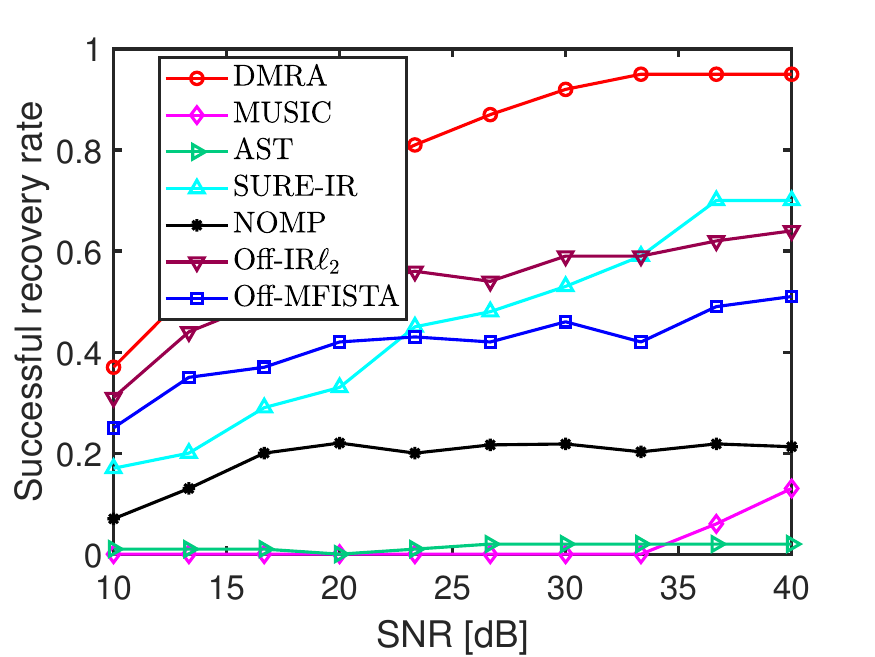}}
    \caption{The RSNR and success rates of various algorithms for different normalized SNR in Scenario 2. The result is the average of 100 independent tests. (a): The average RSNR of different algorithms. (b): The success rates of different algorithms.}
\label{fig:3}
\end{figure}

To better understand the impact of point source structure on super-resolution, we examine Scenario 2, where 8 point sources are distributed in two clusters. Fig.~\ref{fig:3} presents the average RSNR and success rates of different algorithms as a function of normalized SNR. DMRA excels across various SNRs, especially in mid-to-high ranges, with significantly higher RSNR and success rates, indicating excellent reconstruction accuracy and stability. DMRA maintains stable performance across both low and high SNRs, consistently achieving success rates above 0.9 under high SNR conditions. SURE-IR and Off-IR$\ell_2$ show comparable RSNRs to DMRA, particularly in low SNRs. In contrast, AST and MUSIC exhibit near-failure performance in this scenario.

\begin{figure}[!t]
    \centering
    \subfloat[\label{fig:4.1}]{
    \includegraphics[width=2.36in]{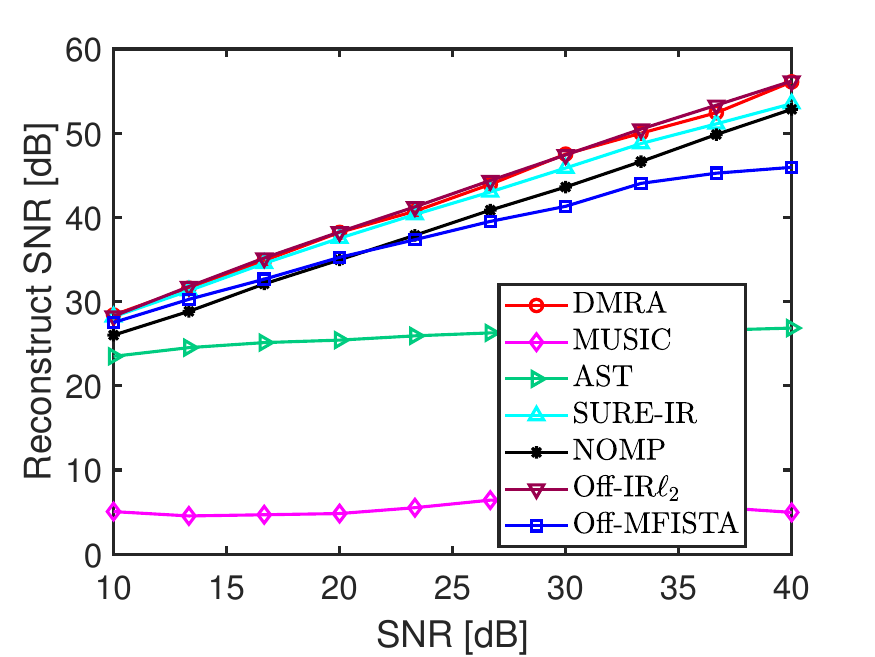}}
    \hfill
    \hspace{0\textwidth}
    \subfloat[\label{fig:4.2}]{
    \includegraphics[width=2.36in]{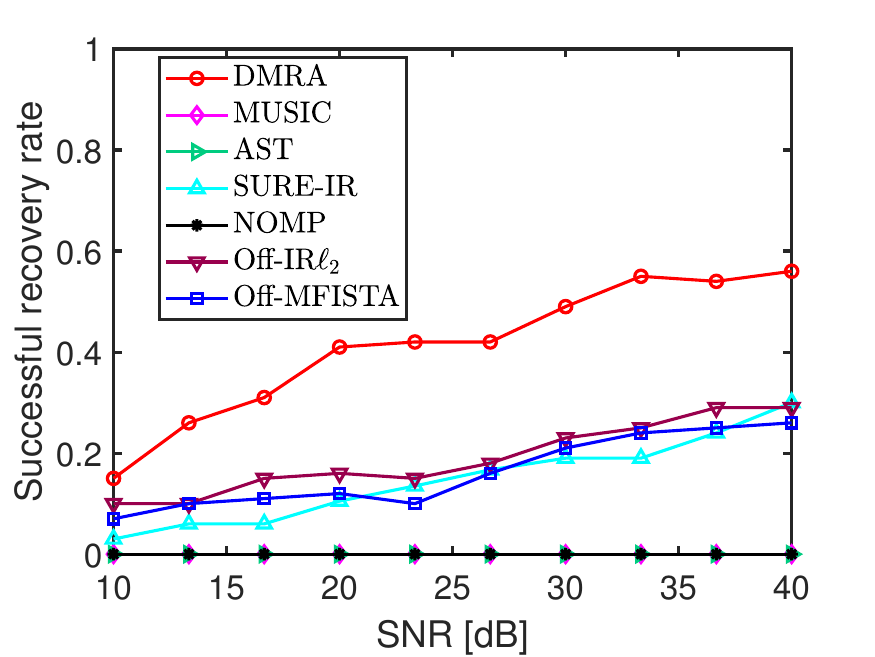}}
    \hfill
    \caption{The RSNR and success rates of various algorithms for different normalized SNR in Scenario 3. The result is the average of 100 independent tests. (a): The average RSNR of different algorithms. (b): The success rates of different algorithms.}
\label{fig:4}
\end{figure}

In Scenario 3, we assess the impact of multiple point sources within a single cluster on super-resolution. Fig.~\ref{fig:4} shows the average RSNR and success rates of different algorithms as a function of normalized SNR. DMRA has a clear advantage in success rates and ties with Off-IR$\ell_2$ in RSNR. NOMP fails in Scenario 3, with success rates near zero, indicating its ineffectiveness for densely packed clusters. Off-MFISTA, SURE-IR, and Off-IR$\ell_2$ have similar success rates, but the RSNR of  Off-MFISTA is slightly lower at high SNR. AST and MUSIC fail to recover the signal in Scenario 3.

In Figs.~\ref{fig:3.2} and \ref{fig:4.2}, the success recovery rates of some algorithms (e.g., MUSIC, AST and NOMP) are nearly zero, primarily because simulation Scenarios 2 and 3 are more challenging and severe compared to Scenario 1. This observation is consistent with the conclusions of super-resolution theory \cite{Batenkov2021, Li2021}, which indicates that as the number of point sources within a cluster increases, the theoretical super-resolution error grows exponentially.

\begin{figure*}[!t]
    \centering
    \subfloat[\label{fig:5.1}]{%
    \includegraphics[width=0.32\linewidth]{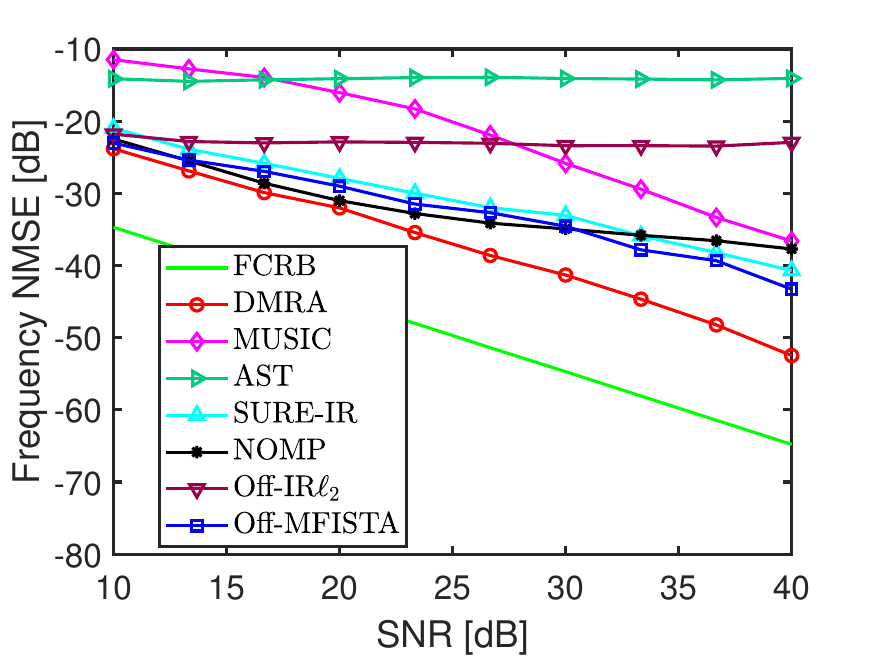}}
    \hfill
    \subfloat[\label{fig:5.2}]{%
    \includegraphics[width=0.32\linewidth]{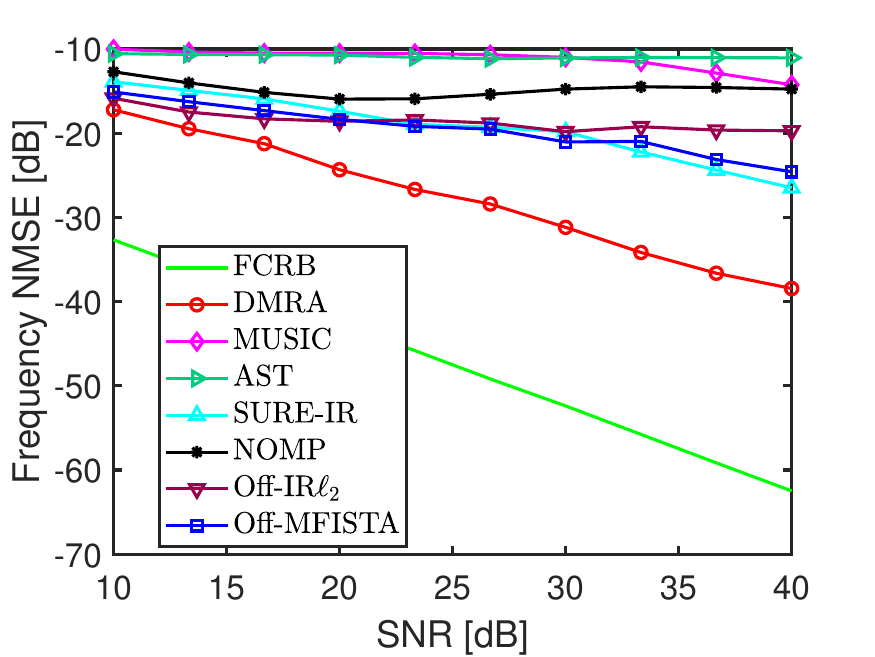}}
    \hfill
    \subfloat[\label{fig:5.3}]{%
    \includegraphics[width=0.32\linewidth]{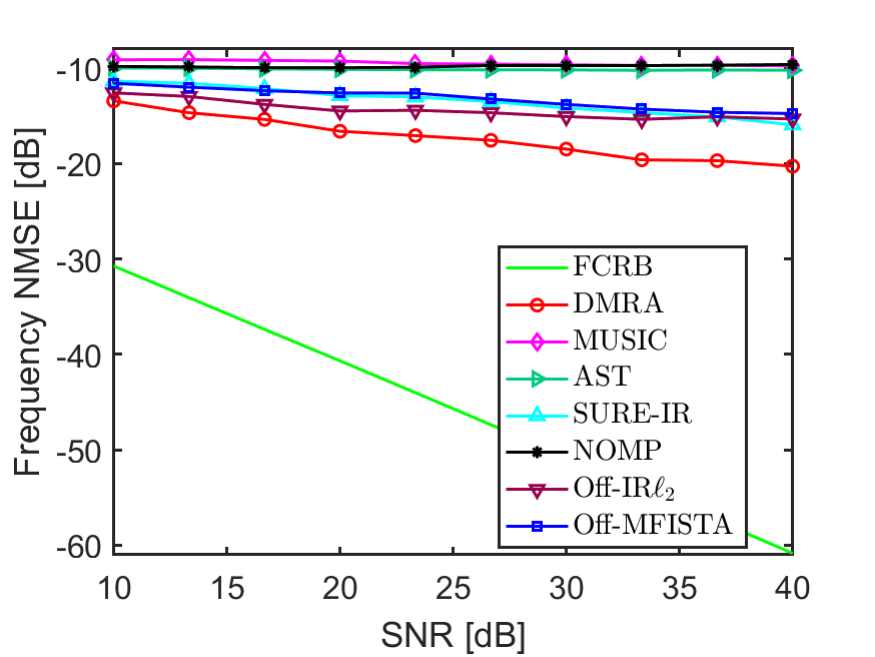}}
    \caption{Frequency estimation error for each algorithm in Scenarios 1-3 as a function of normalized SNR. (a) Frequency estimation errors of algorithms in Scenario 1. (b) Frequency estimation errors of algorithms in Scenario 2. (c) Frequency estimation errors of algorithms in Scenario 3.}
    \label{fig:5}
\end{figure*}

\noindent\paragraph{\textbf{Frequency estimation error}} We compare the super-resolution performance of different algorithms from the perspective of frequency estimation error. The results are shown in Fig.~\ref{fig:5}. The results indicate that DMRA performs excellently across different SNR levels, with more significant advantages in Scenarios 2 and 3. AST performs poorly across all scenarios, while MUSIC performs relatively well in Scenario 1. Off-IR$\ell_2$ shows a saturation effect, where performance stabilizes in the high SNR region, and its frequency estimation error is slightly lower than SURE-IR in low-to-moderate SNRs. Off-MFISTA outperforms SURE-IR in Scenario 1 but degrades significantly in high SNR regions of Scenarios 2 and 3, where SURE-IR performs better. Overall, DMRA significantly outperforms the other algorithms, especially at mid-to-high SNRs. 

We also remark that all tested algorithms perform best in Scenario 1 and worst in Scenario 3, where frequency estimation error deviates most from the Cramér-Rao Lower Bound (CRLB) in all scenarios with 8 point sources (Detailed derivation of the frequency CRLB may be referred to Chapter 3 in \cite{kay1993fundamentals}). This is consistent with the theoretical analysis in \cite{Batenkov2021, Li2021}, indicating that super-resolution error is related to both minimum separation distance and the number of point sources within clusters.

\begin{figure}[!t]
    \centering
    \subfloat[\label{fig:6.1}]{
    \includegraphics[width=2.36in]{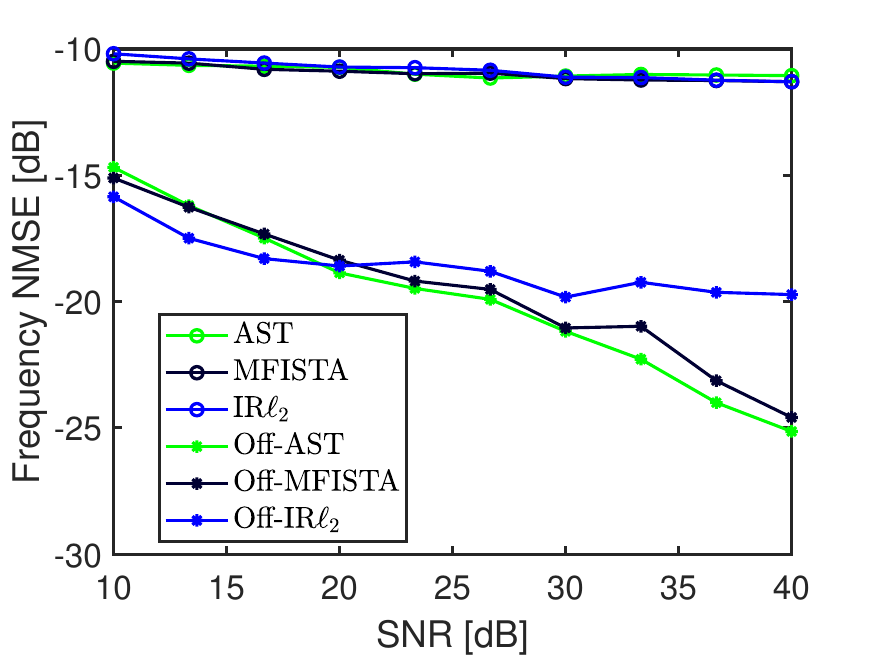}}
    \caption{The frequency estimation error for different algorithms and their off-grid refined version as a function of normalized SNR in Scenario 2.  Off-AST, Off-MFISTA, and Off-IR$\ell_2$ are continuous versions of the original On-Grid algorithms, refined using the BFGS algorithm.}
\label{fig:6}
\end{figure}

It is crucial to note that the off-grid refinement stage significantly enhances the overall algorithm performance. This refinement benefits not only our proposed DMRA method but also other algorithms. For instance, in Scenario 2, we evaluate the impact of off-grid refinement on typical on-grid algorithms: MFISTA, AST, and IR$\ell_2$. Fig.~\ref{fig:6} compares the super-resolution performance of the refined and original algorithms. The refined algorithms show qualitative improvement in super-resolution with only a slight increase in computational cost.

\begin{figure}[!t]
\centering
\subfloat[\label{fig:7.1}]{
\includegraphics[width=2.36in]{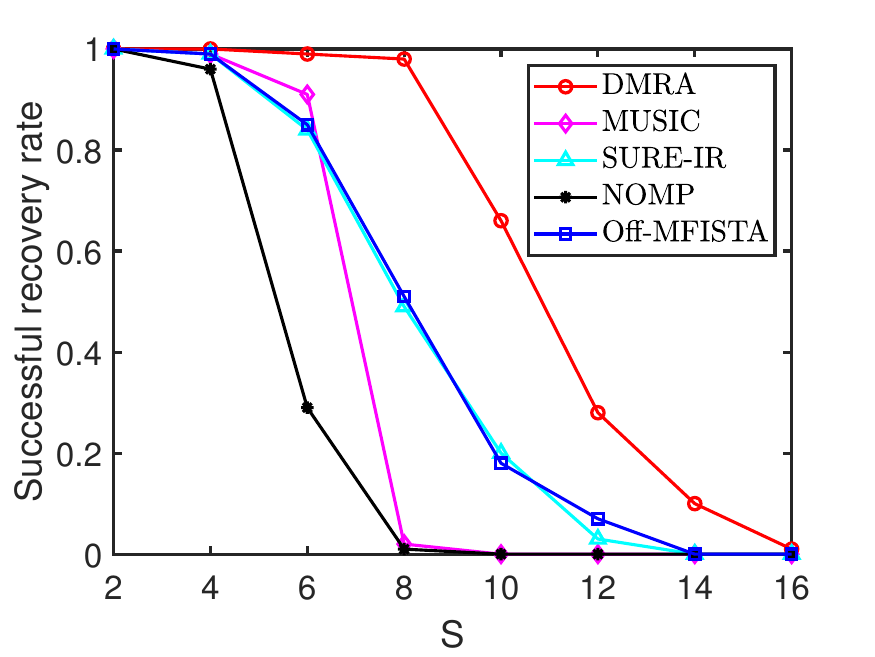}}
\hspace{0.00\textwidth}
\subfloat[\label{fig:7.2}]{
\includegraphics[width=2.36in]{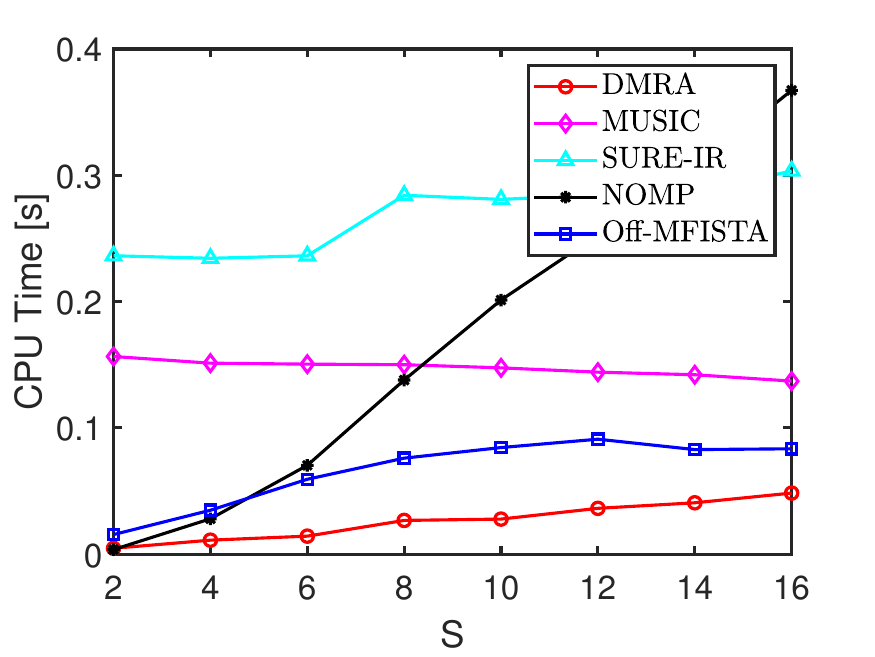}}
\caption{The successful recovery rates and average computation time of various algorithms in Scenario 4. All samples are independently generated at a SNR of 30 dB. The minimum intra-cluster source separation is $\Delta_{\min}(\bm{\omega}) \geq 0.5\Delta_{\mathrm{dft}}$, varying the number of point sources components from $S = 2:2:16$. (a): Successful recovery rates of different algorithms as a function of S. (b): CPU time of different algorithms as a function of S.}
\label{fig:7}
\end{figure}

\noindent\paragraph{\textbf{Computation Time}} In Table ~\ref{tab:2}, we summarize the runtime of each algorithm over 100 simulations in Scenarios 1-3. The DMRA algorithm is the fastest among all algorithms. As the scenario complexity increases, the computation time of DMRA nearly remains constant, whereas the computation time of the other algorithms tends to increase considerably.

\begin{table}[!t]
\centering
\caption{Average Running Times of Respective Algorithms (sec)}
\label{tab:2}
\begin{tabular}{cccc}
    \hline\hline
    Algorithm & Scenario 1 & Scenario 2 & Scenario 3 \\
    \hline
    DMRA & 34.88 & 36.00 & 37.98 \\
    MUSIC & 145.14 & 152.31 & 204.98 \\
    AST & 225.81 & 222.00 & 341.16 \\
    SURE-IR & 509.89 & 475.99 & 577.05 \\
    NOMP & 89.39 & 96.85 & 116.03 \\
    Off-IR$\ell_2$ & 74.34 & 68.45 & 83.33 \\
    Off-MFISTA & 61.47 & 68.51 & 83.63 \\
    \hline\hline
\end{tabular}
\end{table}

\noindent\paragraph{\textbf{Scenario 4}} Fig.~\ref{fig:7} highlights the relationship between super-resolution success rate and the number of discrete point source components for each algorithm. DMRA shows the best overall performance, with high success rates and low computation time, especially as the number of sources increases. The computational complexity of NOMP increases dramatically in dense scenarios. NOMP and Off-MFISTA have low computation time for a small number of sources, but it increases significantly as source numbers grow. Off-MFISTA achieves a success rate comparable to SURE-IR with less computation time. MUSIC maintains relatively stable computation time, but its success rate declines notably as $S$ increases. Overall, DMRA excels in handling noise and sparse signals, making it highly suitable for efficient signal processing in practical applications.

\subsection{5G Channel}

 We also extend our algorithm to a two-dimensional channel scenario. MATLAB 5G Toolbox is used to generate a frequency-domain CDL channel for evaluating channel estimation and extrapolation performance. In the simulation, we consider a 5G CDL channel scenario based on the 3D Urban Macro (3D UMa) model. The carrier frequency is set to 3.5 GHz with a subcarrier spacing of 30 kHz. The total bandwidth is 20 MHz, corresponding to 51 resource blocks (RBs). The simulation involves 2 UE antennas and 64 BS antennas. The channel model follows the CDL-C standard, with a delay spread of 100 ns.

The observations are in the frequency-space domain, assuming that the true frequency-space channel is generated by atoms in the delay-angle domain by $Y = A(\bm{\tau}) X B^{H}(\bm{\theta}) \in \mathbb{C}^{M\times N}$, where $\vec{\tau}$ and $\vec{\theta}$ are delay and angle variables, respectively. $B(\vec{\theta})$ is the dictionary matrix for angle $\vec{\theta}$, which has a similar form as $A$. The matrix $X$ is sparse. In our algorithm, super-resolution recovery is performed in the delay-angle domain. We model $X=\text{diag}\{\vec{x}\}$ as a sparse diagonal matrix and solve the following optimization problem:
\[
\min_{\bm{\tau},\bm{\theta},\vec{x}}\|Y-A(\bm{\tau})\mathrm{diag}\{\bm{x}\}B^H(\bm{\theta})\|_{F}^2 + \lambda \sum_{p=1}^P\mathrm{tanh}\left({|x_p|^2}/{\epsilon}\right).
\]

We employ the DMRA algorithm to solve the super-resolution problem. CDL channel has a typical block sparse structure in the angular domain, where the physical paths within a cluster exhibit close time delays and angles. Due to the presence of a large number of dense paths in the CDL channel, low-energy paths are typically difficult to be resolved by super-resolution techniques. Therefore, by leveraging the block sparsity of the channel and extending the channel parameters based on the super-resolution results obtained from the DMRA algorithm, the accuracy of channel extrapolation can be further enhanced. Denoting the channel parameters estimated by the DMRA method are \((\hat{\bm{\tau}}, \hat{\bm{\theta}})\), and the extended channel parameters are \((\hat{\bm{\tau}}^{e}, \hat{\bm{\theta}}^{e})\). Let \(C_{k} = 2l + 1\) and \(\bm{q} = [-l, -l+1, \cdots, l]^\T/N_t\), we have the following expression:
\begin{equation*}
\hat{\bm{\tau}}^{e} = \mathcal{V}[\hat{\bm{\tau}}^\T\otimes \mathbbm{1}_{C_{k}}], \quad \hat{\bm{\theta}}^{e} = \mathcal{V}[\hat{\bm{\theta}}^\T\otimes \mathbbm{1}_{C_{k}} + \bm{q}\otimes \mathbbm{1}_{C_{k}}^\T].
\end{equation*}

Based on these extended time delay and angle parameters, we estimated the coefficient \(\bm{\hat{x}}^e\) using LMMSE. Subsequently, the CDL-C channel can be extrapolated to \(\hat{Y}^e\) using the following equation:
\begin{equation}
\hat{Y}^e = A(\hat{\bm{\tau}}^e)\mathrm{diag}\{\hat{\bm{x}}^e\}B^H(\hat{\bm{\theta}}^e).
\end{equation}

\begin{figure}[!t]
\centering
\includegraphics[width=2.36in]{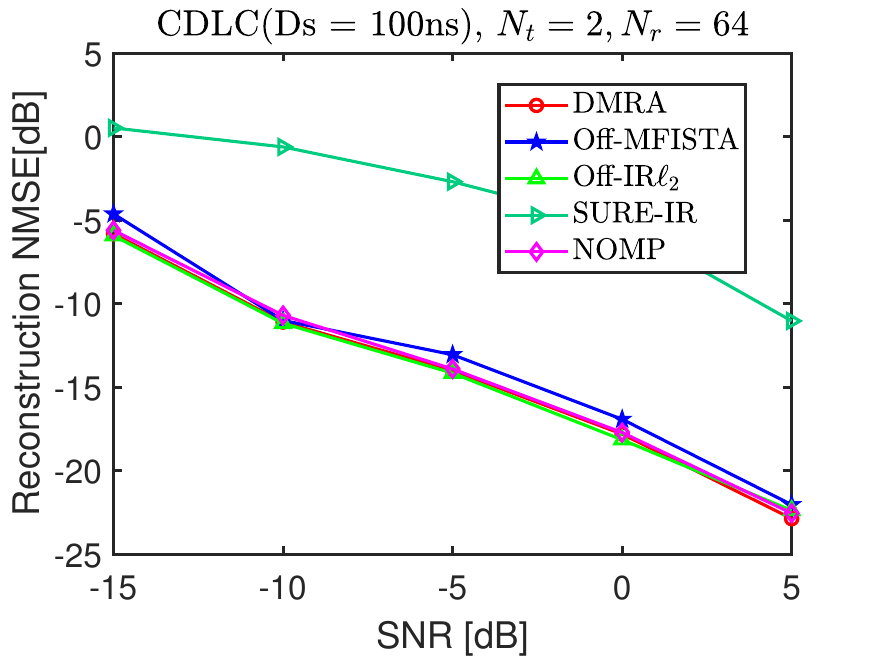}
\caption{Channel reconstruction errors of tested algorithms vs. SNR.}
\label{fig:8}
\end{figure}

\begin{figure}[!t]
\centering
\includegraphics[width=2.36in]{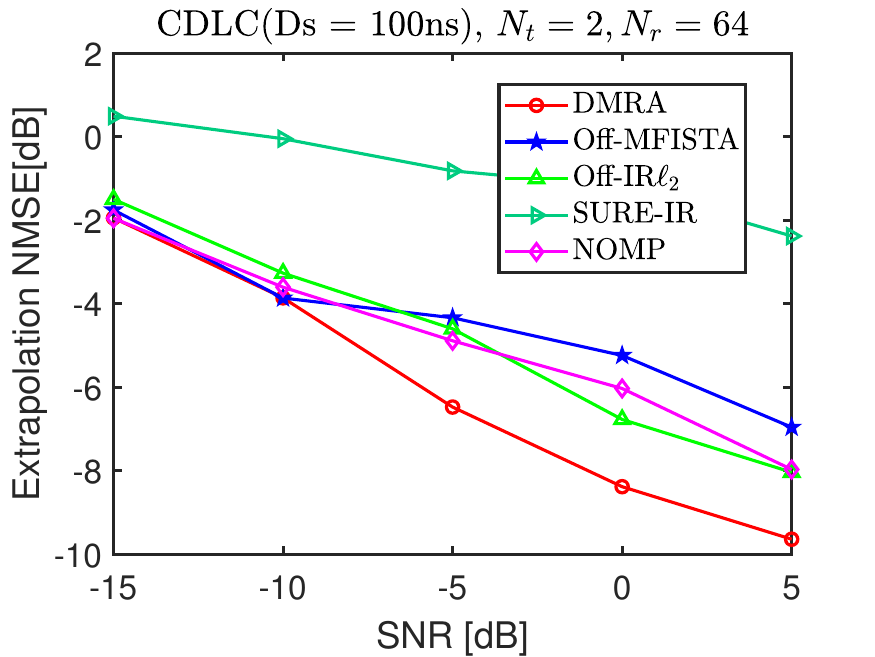}
\caption{Channel extrapolation errors of tested algorithms vs. SNR.}
\label{fig:9}
\end{figure}

\begin{figure}[!t]
\centering
\includegraphics[width=2.36in]{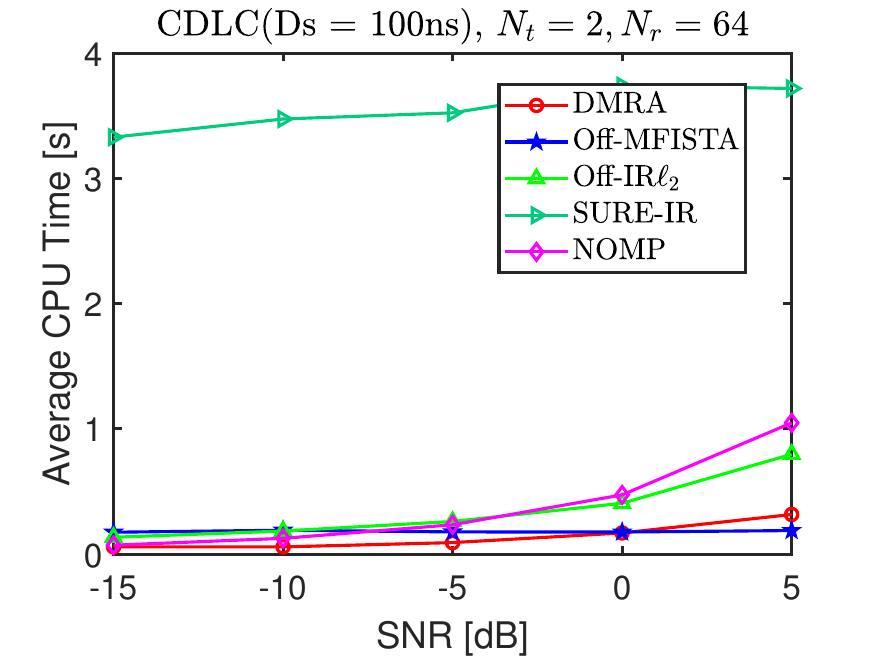}
\caption{Average computation time of tested algorithms vs. SNR}
\label{fig:10}
\end{figure}

In our experiments, we set the grid refinement factor to \(\gamma = 1\) and the parameter controlling the block width to \( l = 10 \). We present the channel reconstruction and extrapolation errors of various algorithms in Figs.~\ref{fig:8} and \ref{fig:9}, respectively. As shown in Fig.~\ref{fig:8}, DMRA, Off-IR\(\ell_2\) and NOMP exhibit similar channel reconstruction errors across various SNRs. SURE-IR has the highest reconstruction NMSE, which performs worse than NOMP.  When the SNR exceeds -10 dB, DMRA demonstrates a significant advantage in extrapolation performance(see Fig.~\ref{fig:9}). Overall, DMRA demonstrates the best performance in channel reconstruction and extrapolation.
In Fig.~\ref{fig:10}, we present the average computation time for different algorithms. It can be observed that the computation time of DMRA increases slowly with increasing SNR, while the computation time of the other algorithms increase more significantly. When $\mathrm{SNR} \geq 0\,\mathrm{dB}$, DMRA has the shortest computation time, followed by Off-MFISTA. At very low SNRs, DMRA and NOMP are both the fastest.

\section{Conclusion}\label{sec:sec6}

We investigate the super-resolution recovery of line spectrum estimation for unknown closely spaced point sources in the continuous domain and proposed an accurate and efficient algorithm DMRA. Consisting of the on-grid and off-grid stages, the proposed DMRA algorithm gradually narrows the subspace, refines and updates the frequency points and the corresponding complex gains through a delicately designed continuous optimization approach. Simulation results demonstrate that the proposed algorithm performs excellently in avoiding energy leakage and exhibits superior super-resolution performance compared to several existing advanced compressed sensing algorithms. Overall, the DMRA algorithm provides a robust and efficient solution for super-resolution line spectrum estimation, which is well-suited for dense signal environments requiring high-resolution parameter estimations.

\section*{Appendix}

\subsection{Proof of Theorem ~\ref{theorem:1}}\label{app:1}

Since the true signal be composed of \( S \) discrete point sources, which cannot be further reduced to fewer components. Therefore, \(\|\tilde{\bm{h}}^r\|_0 = S\). To prove theorem 1, it is sufficient to show that the solution of $A(\vec{\omega})\vec{h}=\vec{y}, s.t. \norm{\vec{h}}_0=S$ is unique. It is easy to see that the true signal $\vec{\omega}^r, \vec{h}^r$ is a solution to this problem. Since $M\geq 2S$, the observed signal \(\bm{y}\) is composed of \( S \) discrete point sources, described by
\begin{equation}\label{eq:22}
y_m  = \sum_{s=1}^S {h}_{s}^r e^{-\j 2\pi (m-1) {\omega}_s^r} = \sum_{s = 1}^S h_{s}^r z_s^{m-1}, S + 1 \leq m \leq 2S,
\end{equation}
where $z_s=e^{-\j 2 \pi \omega_s^r}$. Consider a polynomial whose roots are $z_1,\dots, z_S$:
\begin{equation}\label{eq:23}
\begin{aligned}
P_0(z) := \prod_{s = 1}^S (z - z_s) = \sum_{s = 0}^S \rho_s z^{S-s}.
\end{aligned}
\end{equation}
Denote $\rho_0=1$. The coefficients of the polynomial $\rho_0,\rho_1,\dots, \rho_S$ satisfy the linear system:
\begin{equation}\label{eq:24}
\begin{aligned}
\sum_{s=0}^S  \rho_s y_{m-s}  & = \sum_{s = 0}^S \rho_s \sum_{l=1}^S {h}_l^r z_l^{m-s-1} \\
& = \sum_{l=1}^S {h}_l^r z_l^{m-S-1} \sum_{s = 0}^S \rho_s z_l^{S-s}  = 0,
\end{aligned}
\end{equation}
for $m=S+1,\dots, 2S$.

Define a set of vectors $\bm{y}_k=(y_{k+1},\ldots,y_{k+S})^\T, k=0,1,\dots, S-1$ and matrix $\mathcal{W}=[\vec{y}_{S-1},\dots,\vec{y}_0]$.  The linear system \eqref{eq:24} can be rewrote as $\mathcal{W}\vec{\rho}=-\vec{y}_{S}$.

If the matrix $\mathcal{W}\in\mathbb{C}^{S\times S}$ is non-singular, given observations $\vec{y}$, the coefficients $\vec{\rho}$ is uniquely solvable. So one can solve the polynomial to find the $S$ frequencies $\vec{\omega}^r$. Finally $\vec{h}$ can be uniquely determined by solving the linear system $\vec{y}=A(\vec{\omega}^r)\vec{h}$ since $A(\vec{\omega}^r)$ has full rank. Thus, in order to prove Theorem 1, it is sufficient to show $\mathcal{W}$ is non-singular.

Note \(\bm{y}_k = A(\vec{\omega}^r) D^k {\bm{h}}^r\), where $D=\text{diag}\{e^{\j2\pi {\omega}_1^r},\dots, e^{\j2\pi {\omega}_S^r}\}$. Let \( R^{\mathcal{W}} = \frac{1}{S} \mathcal{W} \mathcal{W}^H \), we have
\begin{equation}\label{eq:26}
\begin{aligned}
R^{\mathcal{W}} & = \frac{1}{S} \sum_{s=1}^S \bm{y}_s \bm{y}_s^H = \frac{1}{S} \sum_{s=1}^S A D^s {\bm{h}}^r (\bm{h}^r)^H (D^s)^H A^H \\
& = A \left(\frac{1}{S} \sum_{s=1}^S D^s {\bm{h}}^r ({\bm{h}}^r)^H (D^s)^H\right) A^H = A R_{\bm{h}} A^H.
\end{aligned}
\end{equation}

According to Theorem 1 in \cite{Shan1985}, the rank of \( R_{\bm{h}} \) is \( S \). Since $A$ is non-singular, $R^{\mathcal{W}}$ is also non-singular. Thus $\mathcal{W}$ is of full rank. The proof is done.

\subsection{Proof of Theorem ~\ref{theorem:2}}\label{app:2}

Assume that $\bm{h}^{\epsilon}$ has $P$ non-zero elements $\{h_p^{\epsilon}\}_{p=1}^P$ with corresponding frequencies $U_P = \{\omega_p^{\epsilon}\}_{p=1}^P$. By Theorem \ref{theorem:1}, $P\geq S$. In case $P = S$, it follows from the proof of Theorem ~\ref{theorem:1} that $\{\bm{\omega}^{\epsilon},\bm{h}^{\epsilon}\} = \{\bm{\omega}^r,\bm{h}^r\}$. Therefore, we only consider the case $S < P \leq N$. Without loss of generality, we assume that $\{h_p^{\epsilon}\}_{p=1}^P$ are sorted in descending order $|h_1^{\epsilon}| \geq |h_2^{\epsilon}| \geq \cdots \geq |h_P^{\epsilon}|$. Under Assumption \ref{assumption:1}, we divide $U_P$ into two subsets: $U_1 = \{\omega_{I_k}^{\epsilon}\}_{k=1}^{J}$ and $U_2 = \{\omega_{I_k}^{\epsilon}\}_{k=J+1}^P$, where $U_1=\vec{\omega}^r\cap U_P$, $U_2=U_P\backslash U_1$, $0 \leq J \leq S$. When $J=0$, $U_1 = \emptyset$. According to Assumption 1, we know that $\forall \omega_i^r \in \vec{\omega}^r, \omega_j^{\epsilon} \in U_2, |\omega_j^{\epsilon} - \omega_i^r| > \tau$, and $\forall \omega_i^{\epsilon}, \omega_j^{\epsilon} \in U_P, |\omega_i^{\epsilon} - \omega_j^{\epsilon}|_{i \neq j} > \tau$, $\forall \omega_i^r, \omega_j^r \in U_{\omega^r}, |\omega_i^r - \omega_j^r|_{i \neq j} > \tau$. These conditions will be used in part (b) of the proof. We denote the index sets of $U_1$ and $U_2$ as $\mathcal{I}^1 = \{I_k\}_{k=1}^J$ and $\mathcal{I}^2 = \{I_k\}_{k=J+1}^P$. The proof will be divided into two parts with part (a): the case $U_1 = \vec{\omega}^r$, $J=S$, and part (b): the general case where $U_1 \subsetneqq \vec{\omega}^r$.

(a) It is easy to see that when $U_1 = \vec{\omega}^r$, $J=S$, according to the properties of the over-determined linear system, $\{h_{I_k}^{\epsilon}\}_{k=1}^{J} = \{h_s^r\}_{s=1}^S$, then
\begin{equation}\label{eq:27}
\begin{aligned}
\mathcal{L}^{\epsilon}(\bm{h}^{\epsilon}) & = \sum_{i \in \mathcal{I}^1} \tanh \left(\frac{|h_i^{\epsilon}|^2}{\epsilon}\right) + \sum_{i \in \mathcal{I}^2} \tanh \left(\frac{|h_i^{\epsilon}|^2}{\epsilon}\right)\\
& \geq \sum_{i \in \mathcal{I}^1} \tanh \left(\frac{|h_i^{\epsilon}|^2}{\epsilon}\right) =\mathcal{L}^{\epsilon}(\bm{h}^r). 
\end{aligned}
\end{equation}

Since $(\bm{\omega}^{\epsilon}, \bm{h}^{\epsilon})$ is the global optimal solution, we have $\mathcal{L}^{\epsilon}(\bm{h}^{\epsilon}) \leq \mathcal{L}^{\epsilon}(\bm{h}^r)$. Thus $\{h_i^{\epsilon}\}_{i=S+1}^P$ are all zero and $\{h_{I_k}^{\epsilon}\}_{k=1}^S = \{h_s^r\}_{s=1}^S$. Hence $\{\omega_{I_k}^{\epsilon}\}_{k=1}^S = \vec{\omega}^r$. So the global optimal solution of $(\mathcal{P}_{\epsilon})$ and $(\mathcal{P}_{0})$ are equivalent.

(b) Now we consider the general case where $U_1 \subsetneqq \vec{\omega}^r$. Let $\mathcal{I}^S = \{p\}_{p=1}^S$ denote the index set of the top $S$ elements of $U_P$, $\mathcal{J}^1 = \mathcal{I}^1 \cap \mathcal{I}^S$, and $\mathcal{J}^2 = \mathcal{I}^2 \cap \mathcal{I}^S$. Let $J^{\prime} = \#|\mathcal{J}^1| = S - \#|\mathcal{J}^2|$. Under these assumptions, we have
\begin{equation}\label{eq:28}
\begin{aligned}
& 0 = \bm{y} - A(\bm{\omega}^r)\bm{h}^r = \sum_{p=1}^{P}h_p^{\epsilon} \bm{a}(\omega_p^{\epsilon}) - \sum_{i=1}^S h_i^r \bm{a}(\omega_i^r)\\
& = \sum_{i \in \mathcal{J}^1}(h_i^{\epsilon} - h_i^r) \bm{a}(\omega_i^r)-\sum_{i \in \mathcal{I}^S \setminus \mathcal{J}^1} h_i^r \bm{a}(\omega_i^r) \\
& + \sum_{i \in \mathcal{J}^2} h_i^{\epsilon} \bm{a}(\omega_i^{\epsilon}) + \sum_{i = S+1}^P h_i^{\epsilon} \bm{a}(\omega_i^{\epsilon})\\
& = \sum_{i \in \mathcal{I}^S} \gamma_i \bm{a}(\omega_i^r) + \sum_{i \in \mathcal{J}^2} h_i^{\epsilon} \bm{a}(\omega_i^{\epsilon}) + \sum_{i = S+1}^P h_i^{\epsilon} \bm{a}(\omega_i^{\epsilon}),
\end{aligned}
\end{equation}
where $\gamma_i := h_i^{\epsilon} - h_i^r$ for $i \in \mathcal{J}^1$ and  $-h_i^r$ otherwise. Let $\gamma_{\max} := \max_{i}|\gamma_i|$. $\mathcal{I}^S \setminus \mathcal{I}^1$ denotes the set obtained by removing the elements common to $\mathcal{I}^S$ and $\mathcal{I}^1$, while retaining the different elements.

If $\gamma_{\max} = 0$, it follows from Assumption \ref{assumption:1} that $\mathcal{I}^1 = \mathcal{I}^S$ and $U_1 = \vec{\omega}^r$. Thus $(\mathcal{P}_{\epsilon})$ and $(\mathcal{P}_{0})$ have the same solution based on similar argument as part (a).

Next we will show that $\gamma_{\max} > 0$ leads to contradiction for $\epsilon$ is small. Moving the first term of Eq.~(\ref{eq:28}) to the LHS and taking norm on both sides, we have:
\begin{equation}\label{eq:30}
\Big\|\sum_{i = S+1}^{P} h_i^{\epsilon} \bm{a}(\omega_i^{\epsilon}) \Big \|_2^2 = \|\Phi(\bm{\omega}) \bm{z}_0\|_2^2\geq \sigma_{\min}^2(\Phi)\norm{\vec{z}_0}_2^2,
\end{equation}
where $\Phi(\bm{\omega}) := [\bm{a}(\omega_1^r), \cdots, \bm{a}(\omega_S^r), \bm{a}(\omega_{\mathcal{J}_1^2}^{\epsilon}), \cdots, \bm{a}(\omega_{\mathcal{J}_{(S-J^{\prime})}^2}^{\epsilon})]$, $\bm{z}_0 := [\gamma_1, \cdots, \gamma_S, h_{\mathcal{J}_1^2}^{\epsilon}, \cdots, h_{\mathcal{J}_{(S-J^{\prime})}^2}^{\epsilon}]$ and $\sigma_{\min}(\Phi)$ is the smallest singular value of $\Phi$. According to Assumption \ref{assumption:1}, the column atoms of $\Phi(\bm{\omega})$ have distinct node frequencies, satisfying $\Delta_{\min}(\bm{\omega}) > \tau$. Thus by Lemma \ref{lemma:1}, we know that
\begin{equation}\label{eq:33}
\|\Phi \bm{z}_0\|_2^2 \geq C^2 M \mu^{4S-2J^{\prime}-2}\norm{\vec{z}_0}_2^2 \geq C^2 M \mu^{4S-2J^{\prime}-2} \gamma_{\max}^2.
\end{equation}

On the other hand,  
\begin{equation}\label{eq:31}
\begin{aligned}
\Big\|\sum_{i = S+1}^{P} & h_i^{\epsilon} \bm{a}(\omega_i^{\epsilon})  \Big\|_2^2  \leq \sum_{i = S+1}^{P} \abs{h_i^{\epsilon}}^2 \|\bm{a}(\omega_i^{\epsilon})\|_2^2\\
& = \sum_{i = S+1}^{P} M|h_i^{\epsilon}|^2  \leq M(P-S) |h_{S+1}^{\epsilon}|^2.
\end{aligned}
\end{equation}
The last inequality holds because $\{|h_i^{\epsilon}|\}_{i=1}^P$ are sorted in descending order. Given that $\mu \geq M \tau$ and $\gamma_{\max} > 0 $, $|h_{S+1}^{\epsilon}|$ can be lower bounded by a positive constant
\begin{align}\label{eq:34}
|h_{S+1}^{\epsilon}|^2 & \geq \frac{C^2 \mu^{4S-2J^{\prime}-2} \gamma_{\max}^2}{P-S}\nonumber\\
 &\geq \frac{C^2 (M\tau)^{4S-2} \gamma_{\max}^2}{P-S} := C_1>0.
\end{align}
The second inequality in Eq.~(\ref{eq:34}) holds because $M \tau < 1$. The lower bound of $\mathcal{L}^{\epsilon}(\bm{h}^{\epsilon})$ can in turn be estimated by
\begin{equation}\label{eq:36}
\mathcal{L}^{\epsilon}(\bm{h}^{\epsilon}) = \sum_{p=1}^{P} \tanh \left({|h_p^{\epsilon}|^2}/{\epsilon}\right)\geq (S+1)\tanh \left({C_1}/{\epsilon}\right).
\end{equation}
The inequality holds because $\{|h_i^{\epsilon}|\}_{i=1}^P$ are sorted in descending order. The lower bound is denoted as $g_{\text{LB}}$.

Meanwhile, we can obtain a upper bound of $\mathcal{L}^{\epsilon}(\bm{h}^\epsilon)$ by 
\begin{equation}\label{eq:37}
\mathcal{L}^{\epsilon}(\bm{h}^{\epsilon})\leq \mathcal{L}^{\epsilon}(\bm{h}^r)= \sum_{n=1}^S \tanh \left({|h_n^r|^2}/{\epsilon}\right)\leq S \tanh \left({|h_{\max}|^2}/{\epsilon}\right).
\end{equation}
The upper bound is denoted as $g_{\text{UB}}$. Let $f(\epsilon) := g_{\text{LB}}(\epsilon) - g_{\text{UB}}(\epsilon)$. 
Clearly, $f(\epsilon)\leq 0$.  However, when $\epsilon < 2C_1/\ln(1+S)$, one may check that $f(\epsilon) > 0$, which is a contradiction. Thus $\gamma_{\text{max}}=0$.

In conclusion, when $\epsilon < 2C_1/\ln(1+S)$, the globally optimal solution of $(\mathcal{P}_{\epsilon})$ is equal to the globally optimal solution of $(\mathcal{P}_{0})$. Combining parts (a) and (b),  we finish the proof of Theorem~\ref{theorem:2}.

\section*{Acknowledgment}

The authors  acknowledge the support from National Key R\&D Program of China under grant 2021YFA1003301, and National Science Foundation of China under grants 12288101 and 12101230.

\end{document}